\documentclass[12pt]{emulateapj} 
\usepackage{amsmath}
\usepackage{amssymb}
\usepackage{amstext}
\usepackage{graphicx}
\usepackage{epstopdf}
\usepackage{epsfig}
\usepackage{color}
\definecolor{myColor}{rgb}{0.9,0.9,0.9}  
\begin{document}
\renewcommand\bottomfraction{.9}
\shorttitle{C/O ratios in exoplanetary atmospheres} 
\title{C/O ratio as a dimension for characterizing exoplanetary atmospheres}
\author{Nikku Madhusudhan\altaffilmark{1}}
\altaffiltext{1}{Department of Physics and Department of Astronomy, Yale University, 
New Haven, CT 06511 {\tt Nikku.Madhusudhan@yale.edu}} 

\begin{abstract}
Until recently, infrared observations of exoplanetary atmospheres have typically been interpreted using models that assumed solar elemental abundances. With the chemical composition fixed, attempts have been made to classify hot Jupiter atmospheres on the basis of stellar irradiation. However, recent observations have revealed deviations from predictions based on such classification schemes, and chemical compositions retrieved from some datasets have also indicated non-solar abundances. The data require a two-dimensional characterization scheme with dependence on both irradiation and chemistry. In hot hydrogen-dominated atmospheres, the C/O ratio critically influences the relative concentrations of several spectroscopically dominant molecules. Between a C/O of 0.5 (solar value) and 2, the H$_2$O and CH$_4$ abundances can vary by several orders of magnitude in the observable atmosphere, and new hydrocarbon species such as HCN and C$_2$H$_2$ become prominent for C/O $\geq$ 1, while the CO abundance remains almost unchanged. Furthermore, a C/O $\geq$ 1 can preclude a strong thermal inversion due to TiO and VO in a hot Jupiter atmosphere, since TiO and VO are naturally under-abundant for C/O $\geq$ 1.  We, therefore, suggest a new two-dimensional classification scheme for hydrogen-dominated exoplanetary atmospheres with irradiation (or temperature) and C/O ratio as the two dimensions. We define four classes in this 2-D space (O1, O2, C1, and C2) with distinct chemical, thermal and spectral properties. Based on the most recent observations, we characterize the thermal structures and C/O ratios of six hot Jupiters (XO-1b, CoRoT-2b, WASP-14b, WASP-19b, WASP-33b, and WASP-12b) in the framework of our proposed 2D classification scheme. While the data for several systems in our sample are consistent with carbon-rich atmospheres (i.e. C/O $\geq$1), new observations are required to conclusively constrain the C/O ratios in the dayside and terminator regions of their atmospheres. We discuss  how observations using existing and forthcoming facilities  can constrain C/O ratios in exoplanetary atmospheres. 
\end{abstract}


\section{Introduction} 
\label{sec:intro}

Recent advancements in observational methods are facilitating detailed characterization of exoplanetary atmospheres. Observations of thermal emission have been reported for over two dozen transiting exoplanets (Seager \& Deming 2010), and have been used to constrain atmospheric temperature profiles (Burrows et al. 2007, Knutson et al. 2008, Madhusudhan \& Seager 2009,2010, Christiansen et al. 2010) and chemical compositions (Fortney et al. 2005, Seager et al. 2005, Barman et al. 2007, Grillmair et al. 2008, Madhusudhan \& Seager 2009,2011, Swain et al. 2009, Stevenson et al. 2010, Madhusudhan et al. 2011) of a substantial number of giant exoplanets. Similar successes, albeit for a smaller set of planets, have also been reported in characterizing exoplanetary atmospheres using transmission spectroscopy (e.g. Charbonneau et al. 2002, Vidal Madjar et al. 2004, Tinetti et al. 2007, Swain et al. 2008, 2012; cf Gibson et al. 2011, Bueaulieu et al. 2008, Desert et al. 2009,  Redfield et al. 2008, Madhusudhan \& Seager 2009, Sing et al. 2011, Bean et al. 2010, Crossfield et al. 2012, Berta et al. 2012), thermal phase curves (Knutson et al. 2007, 2012; Showman et al. 2009; Cowan et al. 2012), and direct imaging (e.g. Marois et al. 2008,2010; Bowler et al. 2010; Currie et al. 2011; Janson et al. 2011). The majority of exoplanets whose atmospheres have been observed, both with ground-based as well as space borne facilities, are highly irradiated giant planets (`hot Jupiters') whose large sizes and high temperatures make their spectra particularly favorable to detect using existing infrared instruments. 

Major efforts have also been directed towards developing a consistent theoretical framework within which to interpret observed spectra of exoplanetary atmospheres. Atmospheric spectra are a function of the temperature structure and the chemical composition of the atmosphere. Early studies typically assumed solar abundances and chemical equilibrium in models of exoplanetary atmospheres and focused primarily on characterizing hot Jupiters based on their temperature structures (Seager \& Sasselov 1998; Sudarsky et al. 2003; Hubeny et al. 2003; Barman et al. 2005; Fortney et al. 2008). Hubeny et al. (2003) demonstrated that highly irradiated atmospheres which can be hot enough to contain gaseous TiO and VO, can host thermal inversions\footnote{A thermal inversions is a region in the atmosphere where temperature increases outward. See e.g. Madhusudhan \& Seager (2010).} in them due to strong absorption of incident visible light by TiO and VO. The presence or absence of a thermal inversion, for a fixed chemical composition, causes distinct spectral signatures which can be identified in observations of thermal emission from hot Jupiters. On this basis, Fortney et al. (2008) proposed two classes of irradiated giant planets based on their incident irradiation. Planets in the hotter class were predicted to host thermal inversions in their atmospheres, caused by the presence of gaseous TiO and VO, whereas those in the cooler class were predicted to lack thermal inversions in their atmospheres due to a lack of TiO and VO; the boundary between the two classes was chosen somewhat arbitrarily at an incident flux of $\sim 10^9$ ergs s$^{-1}$cm$^{-2}$.  

Subsequent observations and theoretical efforts suggested modifications to the TiO/VO hypothesis. While some highly irradiated hot Jupiters do seem to host thermal inversions (Burrows et al. 2007; Knutson et al. 2008; Christiansen et al. 2010; Madhusudhan \& Seager 2010), several counterexamples to the TiO/VO hypothesis have also been reported. For example, some of the least irradiated hot Jupiters are reported as hosting thermal inversions (Machalek et al. 2008,2009), whereas several highly irradiated hot Jupiters show no conclusive evidence for thermal inversions (Fressin et al. 2010; Madhusudhan et al. 2011a; Anderson et al. 2012; Blecic et al. 2012). Spiegel et al. (2009) suggested that TiO and VO, being heavy molecules, may be depleted from the upper atmospheres due to gravitational settling coupled with condensation. Consequently, they suggested, TiO may be responsible for causing thermal inversions in only the most irradiated of hot Jupiters, substantially hotter than the pM-pL boundary of  $\sim 10^9$ ergs s$^{-1}$cm$^{-2}$ originally proposed by Fortney et al. (2008). While the argument  of Spiegel et al could explain the lack of thermal inversions in some hot Jupiters, it does not explain planets with low irradiation which are reported to host thermal inversions (e.g. Machalek et al. 2008). 

More recently, Knutson et al. (2010) reported an empirical correlation suggesting that the chromospheric activity of host stars may inhibit the formation of thermal inversions in hot Jupiter atmospheres by potentially destroying inversion causing absorbers in the planetary atmosphere. While several planets have been know to conform to the correlation, challenges remain. Most of the inferences of thermal inversion considered in the correlation are based on models with fixed solar chemical abundances (e.g. Todorov et al. 2012), while it has been shown that the temperature structure is highly degenerate with chemical composition (Madhusudhan \& Seager 2010). For example, observations of two hot Jupiters TrES-2 and TrES-4 which were originally thought to host thermal inversions in the Knutson et al. study were subsequently found to be consistent with models without thermal inversions (Madhusudhan \& Seager 2010). A similar example of XO-1b is also demonstrated in the present study. Nevertheless, any classification scheme for irradiated giant planets would have to consider the role of stellar activity, as suggested by Knutson et al, in destroying any inversion causing species that may be possible for a given chemistry. 

In view of the above challenges in characterizing hot Jupiter atmospheres, recent observations indicate the requirement of a fundamentally new dimension for characterizing exoplanetary atmospheres -- a dimension which can represent chemistry instead of assuming solar abundances in chemical equilibrium. A key implication from all the previous studies (Hubeny et al. 2003, Fortney et al. 2008, Spiegel et al. 2009, Knutson et al. 2010) is that if the irradiation is high enough and the star is not active, TiO and VO would be prevalent in the atmosphere. Such an implication arises from the fundamental assumption of solar abundances in the atmospheric models on which the above studies based their conclusions. Solar abundances imply an inherently oxygen-rich environment, with a carbon-to-oxygen (C/O) ratio of 0.5, due to which TiO and VO are abundant at high temperatures. However, recent observations suggest that exoplanetary atmospheres may also be carbon-rich, i.e. C/O $\geq$ 1 (Madhusudhan et al. 2011a; Cowan et al. 2012; Swain et al. 2012). For a C/O $\geq$ 1, it has been shown that TiO and VO are naturally under-abundant even at very high temperatures (Helling \& Lucas 2009; Madhusudhan et al. 2011b), thus making any of the other depletion processes such as due to gravitational settling or stellar activity irrelevant. Furthermore, a high C/O ratio fundamentally alters the chemical composition of the atmosphere at high temperatures by depleting strong opacity sources such as H$_2$O and introducing new sources such as CH$_4$, HCN, and C$_2$H$_2$ (Seager et al. 2005; Madhusudhan et al. 2011b; Kopparapu et al. 2012; Moses et al. 2012). Consequently, the C/O ratio can have a substantial influence on the spectroscopic signatures of an exoplanetary atmosphere by influencing both its temperature structure as well as its chemistry. 

Motivated by recent observations, in the present work we suggest a new two-dimensional classification scheme in which the incident irradiation and C/O ratio form the two dimensions. In this 2-D phase space, we identify four classes of H$_2$-dominated atmospheres, namely O1, O2, C1, C2, each with distinct chemical, thermal, and spectroscopic properties. While two of the classes (O1 and O2) pertain to oxygen-rich (C/O $< 1$) chemistry, analogous to the two classes proposed by Fortney et al. (2008), the remaining two classes (C1 and C2) host carbon-rich (C/O $> 1$) chemistry. We discuss in detail the molecular constitution of atmospheres in the different classes and their corresponding spectral signatures, and apply the classification scheme to six hot Jupiters (XO-1b, CoRoT-2b, WASP-14b, WASP-19b, WASP-12b, WASP-33b) with different irradiation levels. We identify a sample of 20 transiting exoplanets which form ideal candidates for characterizing their atmospheric C/O ratios, and populating the 2-D phase space, and for which observations are already available in at least three channels of {\it Spitzer} photometry. By combining the existing observations with new near-infrared observations from ground and space, detailed constraints on the chemistry of atmospheres, interiors, and formation environments, can be obtained for a sizable sample of extrasolar planets. We find in the present work that non-solar C/O ratios can potentially explain observations of hot Jupiters which have hitherto been reported as anomalous using solar abundance models. 

Recently it has become possible to estimate elemental abundance ratios, such as C/H, O/H, and C/O (Madhusudhan et al. 2011a) thus facilitating the thermal and chemical characterization of hot Jupiters in the proposed 2D classification system in the future. Such elemental abundances in planetary atmospheres  also provide critical constraints on the chemistry of their interiors and on their formation conditions. In the solar system, the observed abundances of  volatile elements such as C, S, N, P, and noble gases in Jupiter's atmosphere support its formation by core accretion of planetesimals (Atreya et al. 1999; Atreya \& Wong, 2005) and imply a much cooler origin (at temperature $T < 30$ K) for the planetesimals than Jupiter's present orbit at 5 AU where $T\sim 160$ K (Owen, 1999). One of the most important quantities governing the chemistry in protoplanetary environments is the C/O ratio (Gilman 1969; Larimer 1974; Lodders \& Fegley 1999; Lodders 2004; Kuchner \& Seager 2005; Bond et al. 2010; Madhusudhan et al. 2011b; Oberg et al. 2011). The chemical composition of the planetesimals and gas accreted by the forming planet can be oxygen-rich or carbon-rich depending on whether C/O $< 1$ or C/O $\gtrsim 1$, respectively, in the planet-forming regions. However, the O/H and C/O ratios of solar system giant planets are poorly known since their low atmospheric temperatures ($T \lesssim $ 125 K) cause H$_2$O to condense down to the deeper layers of the atmospheres ($P \sim$ 10 - 100 bar) which are not observable in spectra obtained remotely.  Nevertheless, nominal constraints on the water abundances have been inferred indirectly from CO abundances (Fegley and Prinn 1988, Lodders and Fegley 1994, Bezard et al 2002, Visscher \& Moses 2011). 

The O/H and C/O ratios are easier to measure for atmospheres of highly irradiated extrasolar giant planets 
than they are for solar-system giant planets. The vast majority of exoplanets known have atmospheric temperatures of $T \sim 500 - 3000$ K (e.g. Charbonneau et al. 2005, 2009; Deming et al. 2005; Knutson et al. 2008; Campo et et al. 2009; Smith et al. 2011). At these temperatures, the major carbon and oxygen bearing molecules are in gaseous form in the observable atmosphere and hence can be inferred from spectra (Burrows \& Sharp 1999; Lodders \& Fegley 2002; Madhusudhan \& Seager 2011). The spectroscopically dominant species are typically H$_2$O, CH$_4$, CO, and 
CO$_2$, allowing estimations of C/H, O/H, and C/O ratios. The detectability of these species depends on the spectral resolution and bandpass observed in, and on the molecular abundances, which depend on the atmospheric temperature-pressure profile of a given planet. While a statistically significant constraint (at 3-$\sigma$) on the C/O ratio of an exoplanetary atmosphere has been possible only recently (Madhusudhan et al. 2011a), nominal 
constraints on atmospheric C, O, and/or C/O have been reported for several transiting exoplanets in the past decade (e.g. Fortney et al. 2005; Seager et al. 2005; Madhusudhan \& Seager 2009,2010,2011; Desert  et al. 2009; Swain et al. 2009). However, currently available facilities in space and on ground can be strategically used for high-confidence estimates of C/O ratios of a large sample of transiting exoplanets, as will be shown in the present work. The {\it James Webb Space Telescope (JWST)} will allow estimations of abundance ratios for a larger set of  elements. 

In what follows, we describe our atmospheric modeling methods in Section~\ref{sec:model}. 
We introduce the C/O ratio in Section~\ref{sec:c-o} as a viable new dimension for atmospheric characterization 
of exoplanets. Here, we study in detail the influence of the C/O ratio on the chemical and spectroscopic 
properties of exoplanetary atmospheres and make observable predictions. We introduce a two-dimensional classification scheme for H$_2$-dominated atmospheres in Section~\ref{sec:2d}. In Section~\ref{sec:results}, we 
report model fits to observations of six hot Jupiters, which show preliminary evidence 
for non-solar C/O ratios. In Section~\ref{sec:obs_planning}, we discuss observational approaches 
that might be most beneficial for characterizing exoplanetary atmospheres using existing facilities, and we 
identify currently known transiting exoplanets that are optimal for constraining 
C/O ratios. Finally, we summarize our findings in section~\ref{sec:discussion}. 

\begin{figure*}[ht]
\centering
\includegraphics[width = 0.8\textwidth]{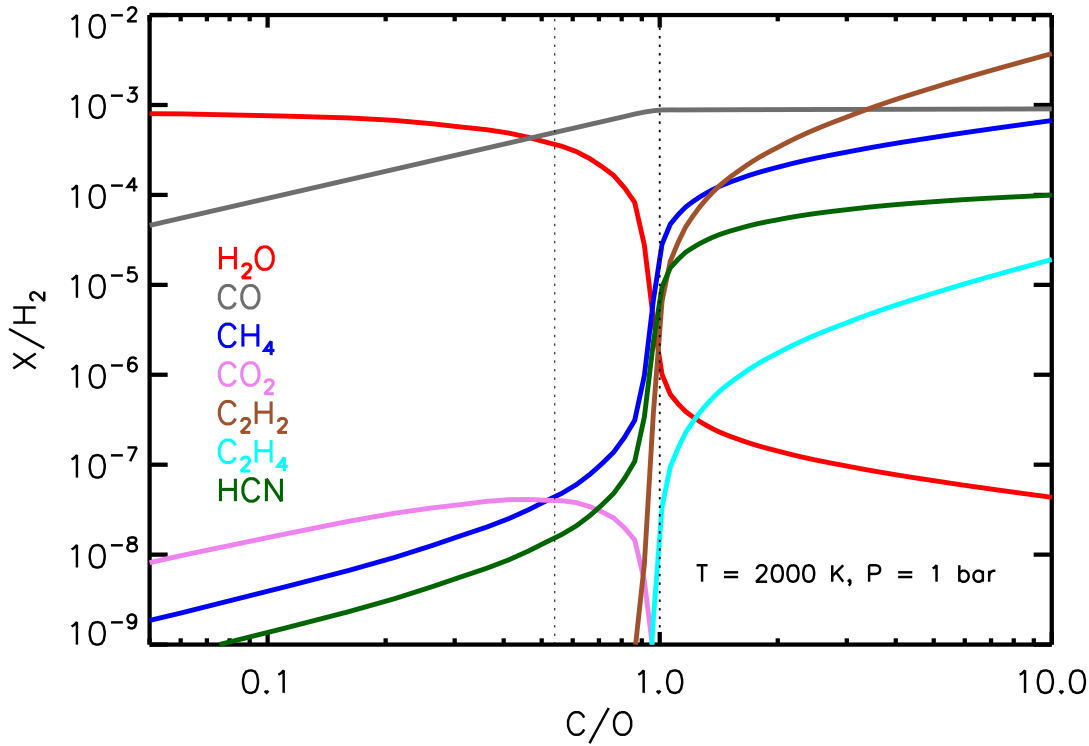}
\caption{Effect of C/O ratio on the volume mixing ratios of major species in H$_2$-dominated atmospheres in chemical equilibrium. The solid curves show mixing ratios, with respect to H$_2$, of H$_2$O, CO, CH$_4$, CO$_2$, C$_2$H$_2$,  C$_2$H$_4$, and HCN, at representative conditions ($T \sim 2000$ K, $P \sim 1$ bar) in observable atmospheres of irradiated giant planets. A sharp transition in the abundances of all the molecules, except CO, occurs at C/O = 1. The dotted vertical line marks C/O = 0.54, corresponding to solar elemental abundances. Compared to solar abundance mixing ratios, H$_2$O is depleted and CH$_4$ is enhanced by over two orders of magnitude each for C/O $\geq 1$.}
\label{fig:cxo}
\end{figure*}

\begin{figure*}[ht]
\centering
\includegraphics[width = \textwidth]{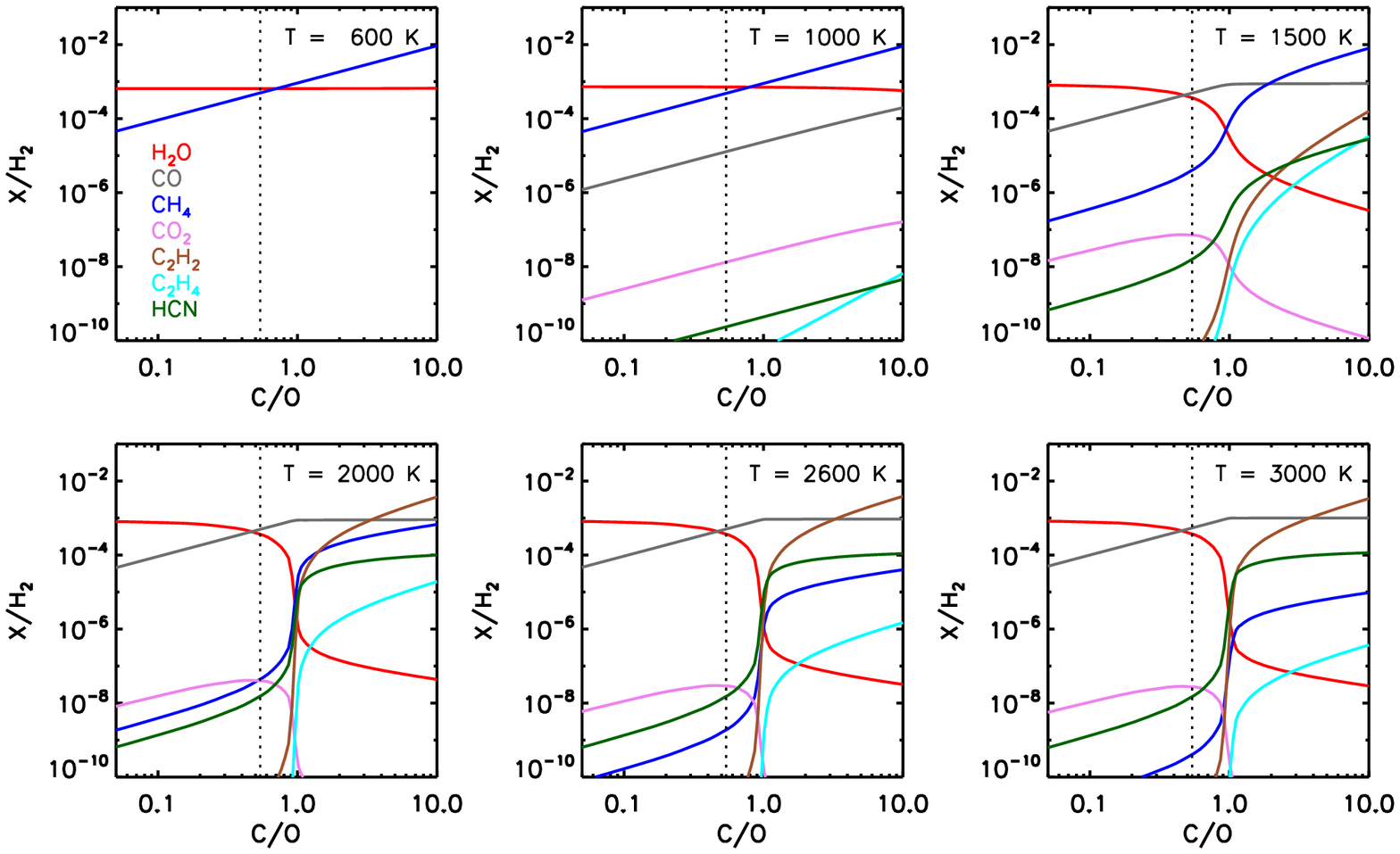}
\caption{Dependance of molecular mixing ratios on C/O ratio and temperature ($T$). Each panel is 
analogous to Fig.~\ref{fig:cxo}, for a particular $T$, shown in the legend, and $P$ = 1 bar. At low $T$ ($\lesssim$ 1200 K), 
H$_2$O is the dominant oxygen carrier and CH$_4$ is the dominant carbon carrier, and CO is least abundant of the three; CO mixing ratio is below $10^{-10}$ for $T \lesssim 600$ K. At high $T$ ($\gtrsim$ 1200 K), the mixing ratios of H$_2$O and CH$_4$, and other hydrocarbons, are critically dependent on the C/O ratio, as discussed in Section~\ref{sec:c-o_1}.} 
\label{fig:cxo_many}
\end{figure*}

\section{Methods}
\label{sec:model}

In this work, we study the dependence of the chemical, thermal, and spectroscopic properties of hydrogen-dominated exoplanetary atmospheres as a function of two key variables - the C/O ratio and the incident irradiation. Our primary focus in on irradiated transiting exoplanets, mainly hot Jupiters but our results are also applicable to hot Neptunes, and hydrogen-dominated super-Earths\footnote{It is presently unknown whether atmospheres of super-Earths are hydrogen-rich or are dominated by heavy molecules, like terrestrial atmospheres; observations have suggested both possibilities (Bean et al. 2010; Croll et al. 2011; Miller-Ricci et al. 2011).} in similar temperature regimes. While we compute models of cloud-free atmospheres, our results on gas phase chemistry and corresponding spectroscopic signatures are also applicable to cloudy atmospheres. Planets at wide orbital separations are expected to be cloudy, and their spectra contain absorption features of species resulting from gas phase chemistry over the cloud tops (e.g. Madhusudhan et al. 2011c). In what follows, we describe our models for computing the atmospheric chemistry and spectra. 

We use the atmospheric modeling and retrieval technique for exoplanetary atmospheres developed by Madhusudhan \& Seager (2009), henceforth MS09, and Madhusudhan et al. 2011a. In order to compute spectra,  the model treats 1-D line-by-line radiative transfer in a plane-parallel atmosphere, with constraints of local thermodynamic equilibrium (LTE), hydrostatic equilibrium, and global energy balance, and includes the major molecular and continuum opacity sources. 
Since we are concerned with H$_2$ dominated atmospheres, our primary sources of opacity in the infrared 
include, depending on the temperature and C/O ratio, H$_2$O, CH$_4$, CO, CO$_2$, C$_2$H$_2$, HCN, 
NH$_3$, and H$_2$-H$_2$ CIA absorption. We also include TiO and VO in regions of the atmosphere where the 
temperatures exceed the condensation temperatures of TiO and VO (Spiegel et al. 2009). Our molecular line data are from Freedman et al. 2008, Freedman (personal communication, 2009), Rothman et al. (2005), Karkoschka \& Tomasko (2010), and Karkoschka (personal communication, 2011), and Harris et al. (2008). We obtain the H$_2$-H$_2$ collision-induced opacities from Borysow et al. (1997) and Borysow (2002). Given the  parametric temperature and molecular abundance profiles, the model computes a spectrum for the required geometry. For the majority of the work, we compute emergent spectra of thermal emission from the dayside atmospheres as observed at secondary eclipse, though in some instances we also compute transmission spectra as are observed at primary eclipse. 

We use our spectral model in two configurations. While computing sample spectra for representative thermal and chemical profiles we use our model in a forward modeling sense. On the other hand, we also use our models to derive best-fit solutions from observations of thermal emission from several hot Jupiters. In the later case, we use our model as a retrieval method (see Madhusudhan et al. 2011a). In this mode, the goal is to formally retrieve molecular abundances and temperature structures from given datasets, as implemented in Madhusudhan et al. (2011a) to estimate posterior probability distributions for the model parameters. This approach involves no assumptions about the chemical abundances\footnote{This approach is different from other forward models which typically assume solar abundances for exoplanetary atmospheres (e.g. Seager et al. 2005; Burrows et al. 2008; Fortney et al. 2008). Once our solutions are retrieved from the data, they are tested for physical plausibility using chemical equilibrium and non-equilibrium calculations and any deviations are reported.} or layer-by-layer radiative equilibrium, and has been successfully used to constrain non-equilibrium chemistry and non-solar abundances (e.g. MS09, 2011; Stevenson et al. 2010; Madhusudhan et al. 2011a), and temperature inversions (MS09 \& 2010; Christiansen et al. 2010) in several exoplanetary atmospheres. In order to sample the model parameter space, we use a Markov chain Monte Carlo (MCMC) algorithm with a Gibbs sampler (see Madhusudhan \& Seager, 2010; Madhusudhan et al. 2011a). Given the limited amount of observations available for each system, our goal in general is not to determine a unique best-fit model, but instead rule out regions of parameter space that are excluded by the data. 

We place constraints on the temperature structure and chemical compositions of six hot Jupiters with different levels of incident irradiation: XO-1b, CoRoT-2b, WASP-14b, WASP-19b, WASP-12b, and WASP-33b. In each case, we explore models with and without thermal inversions and with different C/O ratios. For each system, we identify sample models with different C/O ratios and/or thermal inversions that can explain the data in a self-consistent manner. And based on the 
model fits to the data we discuss new observations that might help further constrain the C/O ratio and thermal structure of each planet. 

We also investigate in detail the observable effects of the C/O ratio on the major molecular species. Similar studies have been reported in the past by Kuchner \& Seager (2005), Seager et al. (2005), and Madhusudhan et al. (2011b). For given elemental abundances, with a prescribed C/O ratio, and thermodynamic conditions ($P$ and $T$) we compute the concentrations of the major species of carbon and oxygen in chemical equilibrium. For this purpose, we use the equilibrium chemistry code originally developed by Seager et al. (2000), and subsequently used in several recent works (Seager et al. 2005; Miller-Ricci et al. 2009; Madhusudhan \& Seager 2011). The code calculates gas phase molecular mixing ratios for 172 molecules, resulting from abundances of 23 atomic species, by minimizing the net Gibbs free energy of the System. The multi-dimensional Newton-Raphson method described in White et al. (1958) is used for the minimization. The free energies of the molecules are calculated using polynomial fits based on Sharp \& Huebner (1990). 

We study the influence of C/O ratio on all the spectroscopically dominant species containing C and O in hydrogen-dominated atmospheres, in chemical equilibrium. Depending on the C/O ratio those species include H$_2$O, CH$_4$, CO, CO$_2$, C$_2$H$_2$, and HCN. Chemical equilibrium is a valid assumption for the highly irradiated atmospheres we consider in this study (e.g. Moses et al. 2011). However, for cooler planets non-equilibrium chemistry in the form of vertical eddy mixing and photochemistry can cause significant changes in the molecular mixing ratios, and hence, affect their spectral signatures (Zahnle et al. 2009; Line et al. 2010; Madhusudhan \& Seager 2011; Moses et al. 2011; Vischer \& Moses 2011). We consider the effect of C/O ratio on non-equilibrium chemistry of hot Jupiter atmospheres in a companion study (Moses et al. 2012). 

\section{C/O ratio as a dimension for atmospheric characterization} 
\label{sec:c-o}

In this section, we study the influence of the C/O ratio on the atmospheric properties of 
cloud-free hydrogen-dominated atmospheres. Spectra of such atmospheres are dominated by 
features of oxygen and carbon based molecules, O and C being cosmically the most abundant 
elements after H and He. For a given pressure-temperature ($P$-$T$) profile, the C/O ratio is 
the most critical parameter that governs the concentrations of the dominant O and C bearing 
species, and, therefore, affects their spectral features and their influences on thermal profiles. 
Several studies have discussed the effect of the C/O ratio on these various aspects of 
hydrogen-dominated atmospheres (Kuchner \& Seager 2005; Seager et al. 2005; 
Helling \& Lucas 2009; Madhusudhan et al. 2011b; Kopparapu et al. 2012). Here, we explore 
a wider range of temperatures and C/O ratios and re-evaluate the effects of the C/O ratio 
from the standpoint of key observables. 

\subsection{Effect of C/O on C and O Chemistry}
\label{sec:c-o_1}

At high temperatures, two distinct regimes of atmospheric chemistry can be identified based on the C/O ratio. 
Fig.~\ref{fig:cxo} shows the mixing ratios of major O and C based molecules as a function of the C/O ratio, 
for nominal conditions in the observable atmospheres of hot Jupiters ($T = 2000$~K, $P = 1$ bar). A sharp transition in the abundances of all the molecules, except CO, occurs at C/O = 1. For example, in going from the oxygen-rich solar abundances with C/O = 0.5 to a carbon-rich value of 1.5, the H$_2$O mixing ratio drops by about three orders of magnitude and the CH$_4$ mixing ratio increases by over three orders of magnitude. In addition, several hydrocarbons besides CH$_4$, such as C$_2$H$_2$, C$_2$H$_4$, and HCN, also become markedly abundant for C/O $\geq$ 1, 
whereas CO$_2$ becomes negligible, as shown in Fig.~\ref{fig:cxo}. These extreme changes represent the two distinctly observable differences between oxygen-rich (C/O $<$ 1) and carbon-rich (C/O $>$ 1) atmospheres at high temperatures, which make hot Jupiters ideal laboratories for constraining C/O ratios. On the other hand, at these $T$ and $P$ the CO abundance for C/O between 0.5 and 1.5 remains nearly unchanged in comparison. Consequently, the CO/H$_2$O ratio, which is nearly unity for C/O = 0.5, can attain values of up to 10 -- 10$^4$ for higher C/O ratios. 

The influence of C/O on the molecular abundances is temperature-dependent, as shown in Fig.~\ref{fig:cxo_many}. 
The mixing ratios of the major species H$_2$O, CO, and CH$_4$ in a hydrogen-dominated atmosphere are 
governed by the following summary reaction:  
\begin{equation}
\mathrm{CH_4 + H_2O \rightleftharpoons CO + 3 H_2.}
\label{eq:ch4-co}
\end{equation}

At a nominal pressure ($P$) of $\sim$1~bar, reaction (\ref{eq:ch4-co}) favors the conversion of 
CH$_4$ to CO at high $T$ ($\gtrsim$ 1300 K), and the reverse reaction is favored at lower $T$. 
Consequently, CH$_4$ is the dominant carbon-bearing molecule at low $T$ and CO is the 
dominant carbon-bearing molecule at high $T$. In an atmosphere with a solar abundance composition 
(C/O = 0.5) which has surplus oxygen, most of the oxygen is contained in H$_2$O at low $T$. 
At high T, about half the available oxygen pairs up with the carbon atoms in CO, and the 
remaining half is present in H$_2$O. Therefore, H$_2$O is abundant at all $T$ for a solar abundance 
atmosphere.  On the other hand, for C/O $\geq 1$, H$_2$O is still the dominant oxygen carrier at low $T$ 
($\lesssim 1200$ K) where CO is minimal, but at high $T$ where CO is dominant, there is little excess oxygen available to form H$_2$O, contrary to the solar abundance case. The excess carbon available for the C/O $> 1$ 
case is present in the form of abundant CH$_4$ and higher hydrocarbons such as C$_2$H$_2$, C$_2$H$_4$, 
and HCN. 

\begin{figure}[ht]
\centering
\includegraphics[width = 0.5\textwidth]{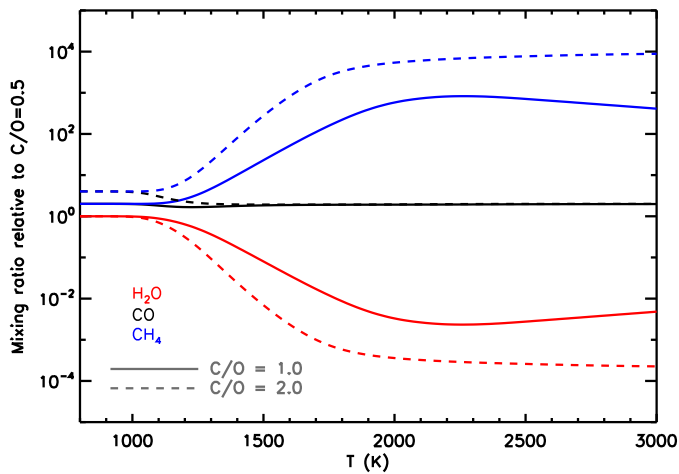}
\caption{Mixing ratios for carbon-rich chemistry relative to solar abundance chemistry as a function of temperature.
The solid (dashed) curves show mixing ratios of H$_2$O, CH$_4$, and CO for C/O = 1 (C/O = 2) relative to those 
obtained for C/O = 0.5. The enhancement of CH$_4$ and depletion of H$_2$O are apparent for $T \gtrsim 1200$ K, 
and can be up to three orders of magnitude for C/O between 1 and 2, and for high $T$. The absolute mixing ratios are shown in \ref{fig:cxo_T2}}
\label{fig:cxo_T1}
\end{figure}

\begin{figure}[ht]
\centering
\includegraphics[width = 0.5\textwidth]{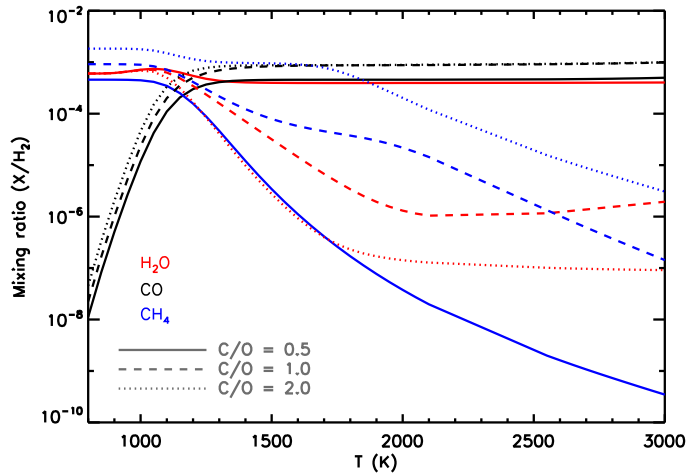}
\caption{Mixing ratios of H$_2$O, CH$_4$, and CO for the cases shown in Fig. \ref{fig:cxo_T1}. Mixing ratios are shown for C/O ratios of 0.5 (solar), 1.0 and 2.0. H$_2$O mixing ratios as low as 10$^{-6}$ and CH$_4$ mixing ratios as high as 10$^{-5}$ can be attained for C/O = 1 and 2, for $T \gtrsim 2000$ K in the lower atmospheres ($P \sim 1$ bar).} 
\label{fig:cxo_T2}
\end{figure}

The relative changes in the H$_2$O and CH$_4$ abundances between the oxygen-rich and carbon-rich 
regimes as a function of temperature are shown in Fig.~\ref{fig:cxo_T1}.  The figure shows enhancements in 
CH$_4$ and depletion in H$_2$O for C/O = 1 and C/O = 2, compared to C/O = 0.5 (solar value); the absolute 
mixing ratios for the different C/O ratios are shown in  Fig.~\ref{fig:cxo_T2}. For $T \lesssim 1200$ K at $P \sim 1$ bar, CO is less abundant, and CH$_4$ is the dominant carbon carrier and H$_2$O the dominant oxygen carrier, for all C/O ratios. In such atmospheres, estimating the C/O ratio requires high precision, to within a factor of 2, estimation of H$_2$O and CH$_4$ abundances; currently molecular abundances for exoplanetary atmospheres are measured to only within a factor of 10 at best (Madhusudhan \& Seager 2009; Madhusudhan et al. 2011a). On the other hand, for $T \gtrsim 1200$ K at $P \sim 1$ bar, as is the case for the majority of transiting exoplanets known, the CH$_4$ enhancement and H$_2$O depletion increase with temperature, and can be over two (three) orders of magnitude for C/O = 1 (C/O = 2) and $T \gtrsim 2000$. In absolute terms, as shown in Fig.~\ref{fig:cxo_T2}, for C/O $\ge1$ the CH$_4$ volume mixing ratios can be as high as $\sim 10^{-5}$ even at $T \gtrsim 2500$ K, contrary to values $\lesssim 10^{-8}$ predicted by a solar C/O ratio. At the same temperatures, for C/O $\geq$ 1, the H$_2$O mixing ratios can be $\lesssim 10^{-6}$, while solar abundances predict $\sim 5 \times 10^{-4}$. These CH$_4$ and H$_2$O mixing ratios for C/O $\geq 1$ at high $T$ are in agreement with the observational constraints for WASP-12b reported by Madhusudhan et al. (2011a). Additionally, other hydrocarbons such as C$_2$H$_2$, C$_2$H$_4$, and HCN, which had not been considered by Madhusudhan et al. (2011a), are also expected to be present in the atmosphere of WASP-12b, as has also been suggested by previous studies (Madhusudhan et al. 2011b; Kopparapu et al. 2012). In Section~\ref{sec:wasp12b}, we demonstrate that model spectra which include these species are consistent with spectroscopic observations of WASP-12b. 

\subsection{Effect of C/O on Thermal Inversions}
\label{sec:c-o_2}

The C/O ratio places important constraints on the likelihood of thermal inversions in 
exoplanetary atmospheres, as discussed in detail in Madhusudhan et al. (2011b). Here, we briefly 
summarize those results. Thermal inversions are caused by the presence of chemical 
species in the observable atmosphere which are strong absorbers of incident UV and visible 
light. Traditionally, TiO and VO have been proposed to cause thermal inversions in highly 
irradiated hot Jupiter atmospheres (Hubeny et al. 2003; Fortney et al. 2008, but cf Spiegel et al. 2009). 
However, Madhusudhan et al. (2011b) show that for C/O $\geq 1$, the TiO and VO abundances in 
chemical equilibrium are lower by about two orders of magnitude compared to their abundances for 
a solar C/O ratio which was assumed in previous models of thermal inversions (Burrows et al. 2008; 
Fortney et al. 2008; Spiegel et al. 2009). Non-equilibrium effects such as gravitational settling 
can further deplete the abundances (Spiegel et al. 2009). Additionally, stellar activity might also effect the 
photochemical stability of inversion-causing compounds (Knutson et al. 2010). Consequently, if  C/O $\geq 1$, even the most highly 
irradiated atmospheres cannot host thermal inversions due to TiO/VO, which might explain the lack of 
a thermal inversion in WASP-12b (Madhusudhan et al. 2011a,b). Conversely, very highly irradiated atmospheres which do not host thermal inversions may likely be carbon-rich. 

C/O ratios derived from observations can help constrain the chemical nature of inversion-causing absorbers. Besides TiO and VO, other oxygen bearing molecules are also unlikely to cause thermal inversions in C-rich atmospheres. For C/O $\geq 1$, almost all the oxygen is occupied by CO, which is the dominant O bearing molecule at high $T$, leaving little O for other oxygen-bearing molecules. As such, if thermal inversions are detected in high C/O atmospheres, they must be formed either by hydrocarbons or by other oxygen-free compounds, e.g. sulfur species (Zahnle et al. 2009). Hydrocarbon-based photochemical products are responsible for thermal inversions in giant planets in the solar system, which are at much lower temperatures ($T \sim 50 - 150$) compared to hot Jupiters  (Chamberlain 1978; Yung \& Demore 1999). However, whether such absorbers can sustain in highly irradiated hot Jupiters ($T \sim 1000 - 3000$) remains to be investigated in future studies. Thus, a simultaneous constraint on the C/O ratio and the presence of a thermal inversion from a given data set is essential to characterize the chemistry of inversion causing absorbers in irradiated giant planet atmospheres. 

\subsection{Characteristic $P$ and $T$}
\label{sec:quench}
The measurability of the C/O ratio in an exoplanetary atmosphere depends on the 
characteristic temperature of the atmosphere. As shown in Figs.~\ref{fig:cxo_many} and \ref{fig:cxo_T1}, 
the variations in the H$_2$O and CH$_4$ mixing ratios with the C/O ratio are significant 
only for $T \gtrsim 1200 K$. Above 1200 K, the differences in the H$_2$O and CH$_4$ abundances 
between the O-rich and C-rich regimes increase with $T$, reaching factors as high as $10^3$ (see Fig. \ref{fig:cxo_T1}). Consequently, spectra of hotter atmospheres are generally more likely to carry stronger signatures 
of their C/O ratios, and would hence be more effective in distinguishing between low ($< 1$) and high ($>1$) C/O models. 

The chemistry in the observable atmosphere (at $P \sim$ 0.01 - 1 bar; SI of Madhusudhan et al. 2011a) 
is influenced by the temperature and pressure in the lower layers of the atmosphere. Vertical 
mixing in the atmosphere transports species from hotter lower regions of the atmosphere 
to upper regions, resulting in nearly uniform concentrations in the observable atmosphere 
(Cooper \& Showman 2007; Zahnle et al. 2009; Line et al. 2011; Madhusudhan \& Seager 2011; 
Moses et al. 2011). Therefore, the pressure level from where the species can be transported up 
(called the `quench' pressure, $P_q$), and the temperature ($T_q$) there, are the 
equilibrium conditions governing the observed concentrations of the species. 

While $P_q$ and $T_q$ can be computed for planet-specific models based on non-equilibrium 
chemistry (e.g. Madhusudhan \& Seager 2011; Moses et al. 2011), we adopt generic values for our current 
purpose. For irradiated giant planet atmospheres, $P_q$ typically 
lies in the 1 - 10 bar range (Cooper \& Showman, 2007; Madhusudhan \& Seager 2011; 
Visscher \& Moses 2011). We assume a nominal value of $P_q = 1$ bar; larger $P_q$ 
requires stronger vertical mixing. As for $T_q$, the dayside $P$-$T$ profiles of strongly 
irradiated atmospheres asymptote to isotherms at high optical depth, for 
$P \sim 1 - 10$ bar (Burrows et al. 2008; Hansen 2009; Guillot 2010). The corresponding 
temperature ($T_q$) is a function of the visible and infrared 
opacities, the Bond albedo, and day-night energy redistribution (Hansen 2008; Guillot et al. 2010; Heng et al. 2011), 
and can be quite high (up to a factor of $\sim$ 1.4) compared to the equilibrium temperature which is given by
\begin{equation}
T_{\rm eq} = T_\star[(1-A_B)(1-f_r)(R_\star^2/2a^2)]^{1/4}, 
\end{equation}
where, $T_\star$ and $R_\star$ are the effective temperature and radius of the host star, 
and $a$ is the orbital separation. $A_B$ is the Bond albedo and $f_r$ is the fraction of incident flux redistributed 
to the night side (Madhusudhan \& Seager 2009). Consequently, based on comparisons with 
published profiles of several irradiated atmosphere models (Burrows et al. 2008, Fortney et al. 2008, 
Madhusudhan \& Seager 2009), we find that a lower limit on $T_q$ can be approximated by: 
 \begin{equation}
T_{\rm q} \gtrsim T_\star[R_\star^2/2a^2]^{1/4} = T_{\rm eq}|_{f = 0}, 
\end{equation}
where $f = 0$ stands for $A_B = f_r = 0$.

We note that our estimation of $T_q$ is only nominal, since several factors of non-equilibrium chemistry,  
such as vertical eddy mixing and photochemistry, can play a role in determine the quench levels of the 
species and $T_q$ can be different for different species (Moses et al. 2011). 

The above estimates for $P_q$ and $T_q$ can be used to identify systems that are most suitable for 
C/O measurements. Irradiated atmospheres with $T_q \gtrsim 1200$ K at $P_q = 1$ bar can be 
expected to contain measurable signatures of their C/O ratios, as shown in Figs.~\ref{fig:cxo_many} and 
\ref{fig:cxo_T1}. The signatures are enhanced with temperature, and are maximimum for $T \gtrsim 2000$ K.

\subsection{Effect of C/O on Atmospheric Spectra}
\label{sec:spectra}

The chemistry in C-rich (C/O $\ge$ 1) and O-rich (C/O $<$ 1) atmospheres present distinct spectroscopic 
signatures that are observable with current and forthcoming instruments. Atmospheres of 
transiting planets can be observed in transit spectra as well as in thermal spectra at occultation\footnote{
The spectral signatures of thermal spectra of transiting planets should also apply to spectra 
of directly imaged planets at similar effective temperatures, though the pressure-temperature 
profiles would be different.}. A transit spectrum constrains the chemistry at the limb 
of the planetary atmosphere, whereas a thermal spectrum constrains the chemistry as well 
as the temperature structure in the dayside atmosphere of the planet 
(Madhusudhan \& Seager 2009). 

\begin{figure}[t]
\centering
\includegraphics[width = 0.5\textwidth]{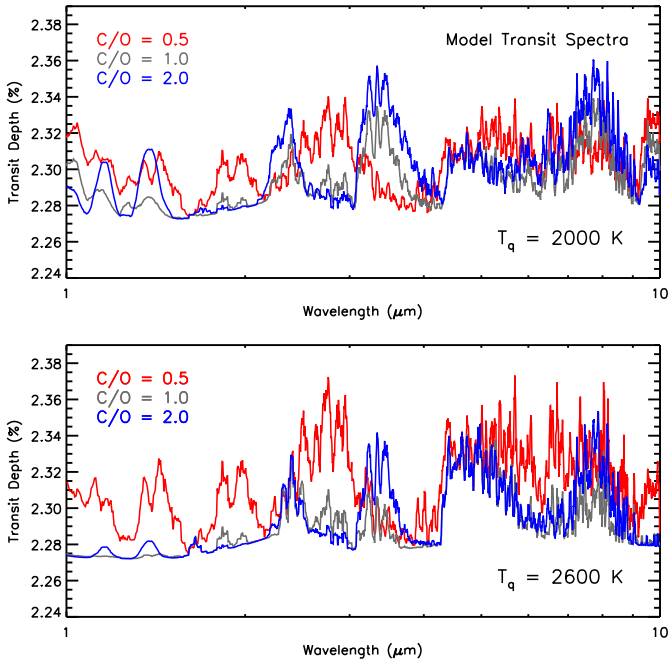}
\caption{Model transit spectra of hydrogen-rich atmospheres with different C/O ratios and temperature structures. 
The planetary properties (radius and surface gravity) are adopted from those of HD~189733b. Spectra are 
shown for C/O ratios of 0.5, 1.0, and 2.0, for two different pressure-temperature ($P$-$T$) profiles. The upper (lower) panel corresponds to a $P$-$T$ profile with a temperature of 2000 (2600) K in the lower atmosphere ($P = 1$ bar). 
In models with C/O = 0.5 (red spectra), the peaks correspond to absorption features of H$_2$O and CO; CH$_4$ is negligible. For C/O = 1 (gray spectra), on the other hand, the H$_2$O features are substantially diminished, and 
mild features of CH$_4$ are visible, along with CO features. For C/O = 2, H$_2$O features are almost non-existent., and strong CH$_4$ features are apparent, in addition to CO features, as discussed in Section~\ref{sec:spectra}.} 
\label{fig:transit}
\end{figure}

\begin{figure}[t]
\centering
\includegraphics[width = 0.5\textwidth]{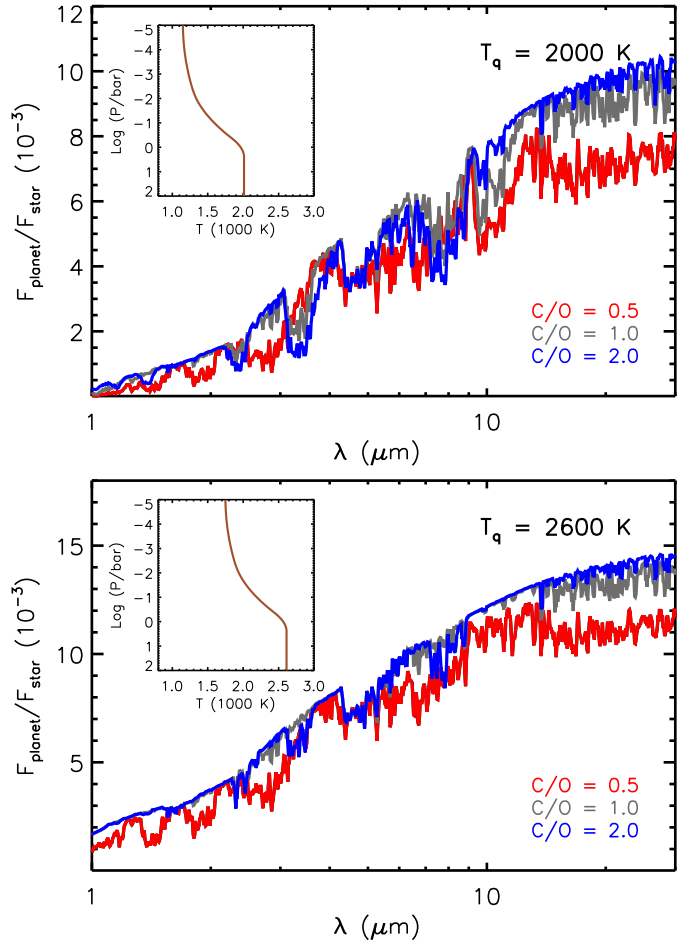}
\caption{Model thermal emission spectra of hydrogen-rich atmospheres with different C/O ratios and temperature structures. Each panel shows spectra for a range of C/O ratios for the pressure-temperature ($P$-$T$) profile shown in the inset. The red, blue and green spectra correspond to C/O ratios of 0.5, 1.0, and 2.0, respectively. The influence 
of C/O ratios on the spectroscopic features are discussed in Section~\ref{sec:spectra}.} 
\label{fig:seclipse}
\end{figure}

In this section, we briefly illustrate the effect on the C/O ratio on the spectra of highly irradiated 
hydrogen-dominated planets. Models for cooler atmospheres have been reported by 
Kuchner \& Seager (2005) and Madhusudhan et al. (2011b). For ease of illustration, we consider 
only four molecules (H$_2$O, CO, CH$_4$, and CO$_2$) in the models discussed here; detailed 
models including additional opacities due to HCN and C$_2$H$_2$ that are possible in C-rich atmospheres 
will be demonstrated in Section~\ref{sec:results}. Model transit and thermal spectra of transiting hot 
Jupiters with different C/O ratios and temperature profiles are shown in Figs.~\ref{fig:transit} \& \ref{fig:seclipse}. 
Here, we use the planetary and stellar parameters of the canonical hot jupiter HD~189733b, and vary the 
temperature profile (adapted from Madhusudhan \& Seager 2009) to span different high-temperature 
regimes, which is equivalent to changing the orbital separation.  

We generate model spectra for two different temperature profiles, shown in the 
insets in Fig. \ref{fig:seclipse}, with different quench temperatures. The quench temperature ($T_q$), discussed in 
Section~\ref{sec:quench}, is the temperature of the lower atmosphere (at $P \sim 1$ bar), which typically governs the chemistry in the observable atmosphere. Spectra are shown for  $T_q = 2000$ K and  $T_q = 2600$ K, and for C/O ratios of 0.5, 1.0, and 2.0, for each $T_q$. For both $T_q$, models with C/O = 0.5 show strong H$_2$O features near 1.1 $\micron$, 1.4  $\micron$, 1.8  $\micron$, and 3 $\micron$, which are clearly distinguishable from the lack of the same features for models with the C/O $\geq$ 1. On the other hand, models with C/O $\geq$ 1 have enhanced CH$_4$ abundances which cause greater absorption in the CH$_4$ bands near 1.2 $\micron$, 1.3 $\micron$, 2.3 $\micron$, 3.3 $\micron$, and 7.8 $\micron$. However, the strength of the CH$_4$ features depend on the temperature and C/O ratio. For $T_q = 2000$, the methane features are only marginally detectable for a model with C/O = 1, but become the dominant spectroscopic features for C/O = 2. On the other hand, for $T_q = 2600$, only the strongest CH$_4$ bands, at 3.3 $\micron$ and 7.8 $\micron$, are prominent for C/O $\geq$ 1. The reason for the relatively weaker CH$_4$ features for $T \sim 2600$, even for C/O $\geq$ 1, is that even though the CH$_4$ enhancement over solar models is as high as for $T \sim 2000$, the absolute CH$_4$ abundances are lower by a factor of $\sim$ 10, as shown in Fig.~\ref{fig:cxo_many}. Finally, all the models show strong CO features at 2.3 $\micron$ and 4.8 $\micron$, since CO is a prominent molecule in both C-rich and O-rich regimes. 

The spectroscopic features discussed above can also be observed in thermal emission spectra 
observed at secondary eclipse. Model spectra of hot Jupiters in thermal emission are shown in 
Fig.~\ref{fig:seclipse}. In the absence of thermal inversions, spectra 
of high temperature O-rich models show strong absorption in the H$_2$O bands and weak 
absorption in CH$_4$ bands, and vice-versa for C-rich models. It must be noted that for thermal spectra 
stronger absorption means lower flux and hence a weaker signal. 

\begin{figure}[ht]
\centering
\includegraphics[width = 0.5\textwidth]{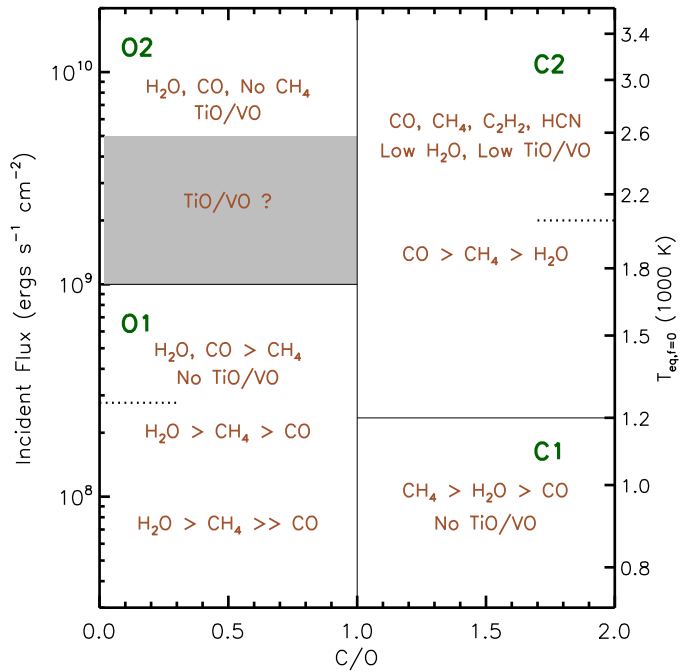}
\caption{A two-dimensional classification scheme for hydrogen-dominated atmospheres 
(see Section~\ref{sec:2d}). The left and right vertical axes show the incident irradiation at 
the substellar point and the equilibrium temperature with no redistribution and no albedo,  
which is representative of the quench temperature (see Section~\ref{sec:quench}). Four classes 
are shown: O1, O2, C1, and C2. Atmospheres in 
O1 and O2 are O-rich (C/O $<1$), and those in C1 and C2 are C-rich (C/O $>$ 1). 
The distinction between O1 (C1) and O2 (C2) is based on irradiation, or temperature. 
The major molecules in each class are shown, for a pressure of 1 bar. Atmospheres in 
O1 and O2 classes are abundant in H$_2$O. CH$_4$ is negligible in O2, but can become 
significant in O1-class atmospheres depending on the temperature; the CO -- CH$_4$ transiting 
temperature is $\sim 1200$ K (see Figs.~\ref{fig:cxo_many} \& \ref{fig:cxo_T1}). Being extremely 
irradiated, atmospheres in the O2 class are more likely to host thermal 
inversions due to gaseous TiO/VO than the O1-class atmospheres; classes O1 and O2 are 
analogous to the pL and pM classes of Fortney et al. (2008). The gray area in the O2 class 
indicates the uncertainty in incident irradiation at which gaseous TiO/VO can remain aloft in 
the upper atmospheres (cf. Spiegel et al. 2009). Atmospheres in C1 and C2 classes are 
under-abundant in H$_2$O compared to their O-rich counterparts at the same irradiation. 
In C2-class atmospheres, H$_2$O is negligible ($\lesssim 10^{-6})$, where as CO, CH$_4$, 
C$_2$H$_2$, and HCN are abundant. In C1-class atmospheres, H$_2$O is significant, 
but still less than their O-rich counterparts. In both C1 and C2 classes, TiO and VO are naturally 
under-abundant even for very high irradiation levels and, hence, cannot cause thermal inversions 
(Madhusudhan et al. 2011b).} 
\label{fig:2d}
\end{figure}

\section{A Two-dimensional Classification Scheme}
\label{sec:2d}

The distinct changes in molecular composition with the C/O ratio and their influence on the observable 
spectroscopic properties make the C/O ratio a viable dimension for characterization of hot exoplanetary atmospheres. Consequently, we suggest a two-dimensional classification scheme for highly irradiated giant planet atmospheres, in 
which incident irradiation and chemistry, represented by the C/O ratio, are the two dimensions. Our approach 
is a generalization of the one-dimensional scheme proposed by Fortney et al. (2008) in which irradiation is the 
only dimension and the composition was fixed to be solar (C/O = 0.54). Our addition of the C/O ratio as an 
extra dimension is also motivated by the possibility that non-solar C/O ratios might explain several of the 
observations which would otherwise be deemed as anomalies, based on the 1-D hypothesis, as shown in Section~\ref{sec:results} and Madhusudhan et al. (2011a).

A schematic representation of the two-dimensional classification scheme is shown in Fig.~\ref{fig:2d}. Hydrogen-dominated irradiated atmospheres can be classified into four classes in this 2-D phase space. Along the C/O axis, C/O = 1 forms a natural boundary between C-rich and O-rich atmospheres with manifestly distinct chemical and spectroscopic characteristics, as shown in Section~\ref{sec:c-o}. Along the irradiation axis, atmospheres can be 
classified broadly in two classes based on the temperatures where significant transitions occur in the chemistry in 
each C/O regime, similar to the 1-D classification scheme of Fortney et al. (2008) for O-rich atmospheres. 
Consequently, we identify the four classes as: 

\begin{itemize}
\item{O1: C/O $<$ 1 and $F \le F_{\rm crit}$}
\item{O2: C/O $<$ 1 and $F \ge F_{\rm crit}$}
\item{C1: C/O $\ge$ 1 and $T_q \le 1200$ K}
\item{C2: C/O $\ge$ 1 and $T_q \ge 1200$ K}
\end{itemize}

In this notation, classes O1 and O2 correspond to oxygen-rich (C/O $<$ 1) atmospheres, and C1 and C2 correspond to carbon-rich (C/O $>$ 1) atmospheres. In each C/O regime, the suffix `1' represents the lower 
temperature (or irradiation) class and `2' represents the higher temperature class. As described in Section~\ref{sec:c-o_1}, for a given irradiation level C-rich atmospheres (in classes C1 or C2) are depleted in H$_2$O and enhanced in CH$_4$, C$_2$H$_2$ and HCN, compared to O-rich atmospheres at the same irradiation (in classes O1 or O2). Within C1 and C2, the amounts of H$_2$O depletion and CH$_4$ enhancement, over solar values, increase with irradiation. In the 
O-rich classes, H$_2$O and CO are abundant in both O1 and O2, except for the lesser irradiated 
atmospheres in O1 (with $T_q \lesssim 1300$ K), where CH$_4$ can be a major carbon carrier and can 
almost entirely replace CO for $T_q \lesssim 1000$ K.  

The division between O1 and O2 along the irradiation axis concerns the possibility of thermal inversions in irradiated atmospheres. O1 and O2 are analogous to the pL and pM classes, respectively, of Fortney et al. (2008). In the O-rich regime, very highly irradiated atmospheres are likely to host gaseous TiO and VO which might cause thermal 
inversions. However, our division of O1 and O2 at an irradiation of $F_{\rm crit} = 10^9$ ergs/s/cm$^2$, following Fortney et al., is only nominal. Spiegel et al (2009) showed that TiO and VO are subject to strong gravitational settling and hence may not be present aloft in the atmosphere in significant quantities for all but the most extremely irradiated systems. Consequently, a more accurate boundary might be at $F_{\rm crit} \sim 5\times 10^9$ ergs/s/cm$^2$. This uncertainty in $F_{\rm crit}$ which is depicted in the gray area, could potentially be constrained by future constraints of thermal inversions in O2 class systems. 

The division between C1 and C2 is motivated by the abundance of H$_2$O and CO. In C-rich atmospheres, TiO and VO are naturally under-abundant, as discussed in Section~\ref{sec:c-o_2}, and hence atmospheres in neither C1 nor C2 can host thermal inversions due to TiO and VO. Consequently, at the current time, there is no motivation to divide the high and low irradiation regimes on the basis of thermal-inversion-causing absorbers. However, two regimes can be defined in C-rich atmospheres based on the effect of temperature on the relative abundances of 
H$_2$O, CO, CH$_4$. As shown in Section~\ref{sec:c-o_1}, the depletion of H$_2$O and enhancement of CH$_4$ in C-rich atmospheres over their O-rich counterparts are substantial only for $T \gtrsim 1200$ K (i.e. in C2 class). Atmospheres in the C2 class would be enriched in hydrocarbons such as CH$_4$, C$_2$H$_2$, and HCN, and be manifestly depleted in H$_2$O. As such, C2 class planets are readily identifiable by their weak H$_2$O absorption bands, and strong hydrocarbon bands. For  $T_q \lesssim 1200$ K, C-rich atmospheres have H$_2$O depletion only by factors of a few compared to O-rich atmospheres with the same irradiation. At such temperatures, C1 class planets may not be readily distinguishable based on H$_2$O abundances alone, unless really precise estimates of the H$_2$O abundances become available. In C1 class objects CH$_4$ is the most abundant molecule, followed by H$_2$O and CO. Additionally, given their low temperatures C1 class objects are most susceptible to non-equilibrium chemistry (Moses et al. 2011), by products of which may aid in their spectroscopic characterization. 

\begin{figure}[ht]
\centering
\includegraphics[width = 0.5\textwidth]{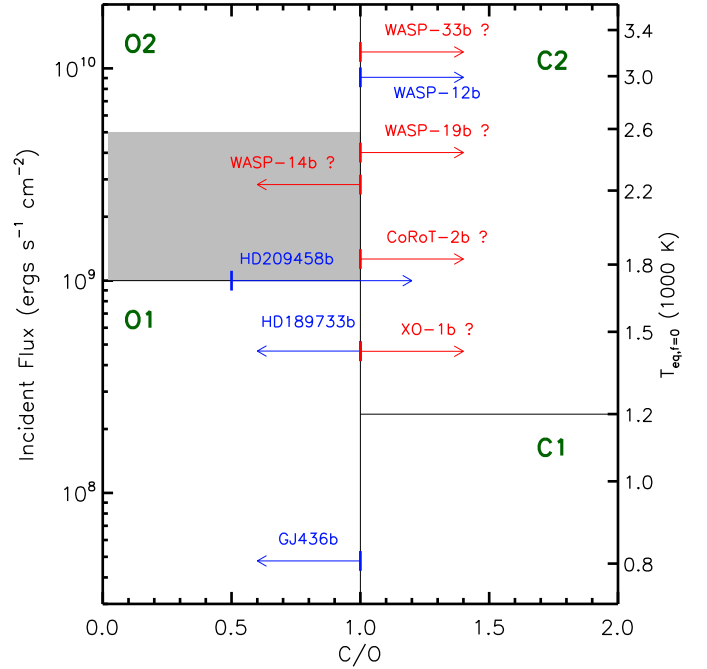}
\caption{Two-dimensional characterization of transiting exoplanets for which constraints on 
C/O ratios are available. See Fig.~\ref{fig:2d} for description of the classification scheme. 
The arrows show the constraints on the C/O ratios. The blue arrows 
show constraints reported in previous studies (see Section~\ref{sec:2d}), and the constraints in red show 
tentative C/O ratios inferred for systems in the present study. A question mark against a planet 
name indicates that its classification is only suggestive and requires new observations for confirmation (see Section~\ref{sec:results}).} 
\label{fig:2d_1}
\end{figure}

Classification of a planet in the present 2-D scheme requires that its irradiation and atmospheric C/O ratio be known. While the incident flux can be obtained from the stellar and orbital parameters, determining the 
C/O ratio requires infrared observations of the planetary atmosphere. As shown in blue in Fig.~\ref{fig:2d_1}, nominal constraints on the C/O ratios, based on at least six channels of broadband photometry and/or spectrophotometry, have been reported for only four transiting planets to date: HD~189733b (Madhusudhan \& Seager, 2009; Swain et al. 2009),  HD~209458b (Madhusudhan \& Seager 2009,2010), GJ~436b (Stevenson et al. 2010; Madhusudhan \& Seager 2011), and WASP-12b (Madhusudhan et al. 2011). Even so, only WASP-12b has a statistically robust constraint of C/O $\geq 1$, and, hence, can be classified as a C2-class system. The C/O ratio of 
GJ~436b is currently uncertain owing to anomalously low CH$_4$ abundances (Stevenson et al. 2010; Madhusudhan \& Seager 2011) which could not be explained by non-equilibrium chemical models to date (Line et al. 2011; Madhusudhan \& Seager 2011). Also shown in Fig.~\ref{fig:2d_1}, in red, are tentative constraints for several hot Jupiters considered in the present study (see Section~\ref{sec:results}).

The 2-D classification scheme presents a testable hypothesis for existing and upcoming observations. 
Since WASP-12b belongs to the C2 class, we expect it to be depleted in TiO and VO. This prediction is 
consistent with the lack of  a noticeable thermal inversion in WASP-12b reported by Madhusudhan et al. (2011). 
We revisit WASP-12b in Section~\ref{sec:results}). Nevertheless, future observations of WASP-12b in transit in the near-infrared (between 0.7 $\micron$ and 1.1 $\micron$) could place constraints on the TiO/VO abundances in its atmosphere (Desert et al. 2008). Similarly, the lack of noticeable thermal inversions in HD 189733b ( Burrows et al. 2008, Charbonneau et al. 2008, Madhusudhan \& Seager 2009) and GJ 436b (Stevenson et al. 2010; Madhusudhan \& Seager 2011) are also consistent with predictions for O1 systems. For the hot Jupiter HD~209458b, the C/O ratio is not well constrained, spanning both O-rich and C-rich regimes, depending on the models used to interpret the data (Seager et al. 2005, Burrows et al. 2008, Madhusudhan \& Seager 2009,2010). Consequently, more data is required to characterize HD~209458b. In the present study, we now consider six hot Jupiters over a wide range of incident irradiation and attempt to characterize their atmospheres in the framework of our 2D classification scheme proposed here. 

\begin{deluxetable*}{l c c c c c c c c c}
\tablewidth{0pt} 
\tabletypesize{\scriptsize}
\tablecaption{Chemical compositions of model spectra}
\tablehead{\colhead{System\tablenotemark{a}} & \colhead{H$_2$O} & \colhead{CO} & \colhead{CH$_4$} & \colhead{CO$_2$} & \colhead{C$_2$H$_2$} & \colhead{HCN} & \colhead{C/O} & \colhead{$\chi^2$}  & \colhead{N$_{\rm obs}$}}
\startdata  
XO-1b \\
ORP & $3\times 10^{-4}$ & $4\times 10^{-4}$ & $1\times 10^{-7}$ & $2\times 10^{-7}$ & $4\times 10^{-10}$ & $2\times 10^{-7}$ & 0.6 & 9.4 & 4 \\
CRP & $1\times 10^{-6}$ & $2\times 10^{-3}$ & $3\times 10^{-5}$ & $4\times 10^{-8}$ & $3\times 10^{-7}$ & $2\times 10^{-6}$ & 1.0 & 5.0 & 4 \\
\hline \\
CoRoT-2b\\
ORP & $2\times 10^{-4}$ & $1\times 10^{-4}$ & $4\times 10^{-9}$ & $7\times 10^{-9}$ & $6\times 10^{-10}$ & $2\times 10^{-8}$ & 0.3 & 5.5 & 4 \\
CRP & $2\times 10^{-6}$ & $2\times 10^{-4}$ & $1\times 10^{-5}$ & $2\times 10^{-10}$ & $4\times 10^{-6}$ & $5\times 10^{-5}$ & 1.3 & 2.9 & 4 \\
\hline \\
WASP-14b\\
ORP & $3\times 10^{-3}$ & $4\times 10^{-4}$ & $1\times 10^{-8}$ & $1\times 10^{-8}$ & $1\times 10^{-8}$ & $1\times 10^{-8}$ & 0.1 & 3.4 & 4 \\
CRP & $1\times 10^{-6}$ & $1\times 10^{-4}$ & $2\times 10^{-6}$ & $6\times 10^{-10}$ & $2\times 10^{-5}$ & $4\times 10^{-5}$ & 1.4 & 5.2 & 4 \\
\hline \\
WASP-19b\\
ORP & $1\times 10^{-3}$ & $6\times 10^{-4}$ & $6\times 10^{-10}$ & $6\times 10^{-9}$ & $2\times 10^{-8}$ & $2\times 10^{-8}$ & 0.4 & 13.4 & 7 \\
CRP & $2\times 10^{-6}$ & $5\times 10^{-4}$ & $5\times 10^{-7}$ & $1\times 10^{-10}$ & $1\times 10^{-5}$ & $1\times 10^{-5}$ & 1.1 & 3.4 & 7 \\
\hline \\
WASP-12b\\
ORP & $1\times 10^{-3}$ & $1\times 10^{-3}$ & $1\times 10^{-9}$ & $1\times 10^{-7}$ & $1\times 10^{-10}$ & $2\times 10^{-9}$ & 0.5 & 48.0 & 8 \\
CRP & $1\times 10^{-6}$ & $2\times 10^{-3}$ & $1\times 10^{-5}$ & $5\times 10^{-9}$ & $1\times 10^{-5}$ & $3\times 10^{-6}$ & 1.0 & 15.8 & 8 \\
\hline \\
WASP-33b\\
ORP & $6\times 10^{-4}$ & $5\times 10^{-4}$ & $1\times 10^{-10}$ & $5\times 10^{-8}$ & $1\times 10^{-10}$ & $1\times 10^{-10}$ & 0.5 & 11.6 & 4 \\
CRP & $2\times 10^{-7}$ & $1\times 10^{-3}$ & $8\times 10^{-5}$ & $9\times 10^{-10}$ & $9\times 10^{-6}$ & $3\times 10^{-6}$ & 1.0 & 4.8 & 4 \\
\enddata
\tablenotetext{a}{The two rows for each system correspond to the two model spectra for each system. 
   ``ORP"  and ``CRP" stand for oxygen-rich planet and carbon-rich planet, respectively. An ORP is defined as a planet with an 
      atmospheric  C/O $<$ 1, and a CRP is defined as a planet with an atmospheric C/O $\geq$ 1 (Madhusudhan et al. 2011b).}
\label{tab:chem}
\end{deluxetable*}

\section{Application to spectral observations of hot Jupiters}
\label{sec:results}

Recent observations of the dayside atmospheres of several hot Jupiters have proved difficult to explain in the 
context of traditional atmospheric models and classification schemes which assume solar 
abundances (i.e. C/O = 0.5). On one hand, several highly irradiated hot Jupiter atmospheres which were previously 
predicted to host thermal inversions due to TiO and VO showed no conclusive evidence of thermal inversions 
in their spectra, e.g. WASP-12b (Madhusudhan et al. 2011), WASP-19b (Anderson et al. 2012), WASP-14b (Blecic et al. 2012). On the other hand, in at least one case a thermal inversion was inferred for one of the least irradiated hot Jupiters, XO-1b (Machalek et al. 2008). In yet another case, no models have thus far been able to explain the 
dayside emission observed for the hot Jupiter CoRoT-2b (Deming et al. 2011). 

In the present section, we demonstrate that all the above mentioned observations can be consistently 
explained in the framework of the 2-D classification scheme proposed in section \ref{sec:2d}, using models 
that do not restrict the atmospheric compositions to solar C/O ratios. We present model 
fits to observations of six hot Jupiters spanning a wide range in stellar irradiation: XO-1b, CoRoT-2b, XO-2b, WASP-14b, WASP-19b, WASP-12b, and WASP-33b. For each planet, we consider models with O-rich (i.e. C/O $<$ 1) as well as C-rich (i.e. C/O $\geq$ 1) compositions, and discuss the degree of fit in each case. Depending on the degree of fit, we  attempt to characterize each planet based on its incident irradiation and possible C/O ratio. Two sample spectra, one O-rich and one C-rich, fitting the data are shown for each planet. The corresponding chemical compositions are shown in Table~\ref{tab:chem}.

\begin{figure}
\centering
\includegraphics[width = 0.5\textwidth]{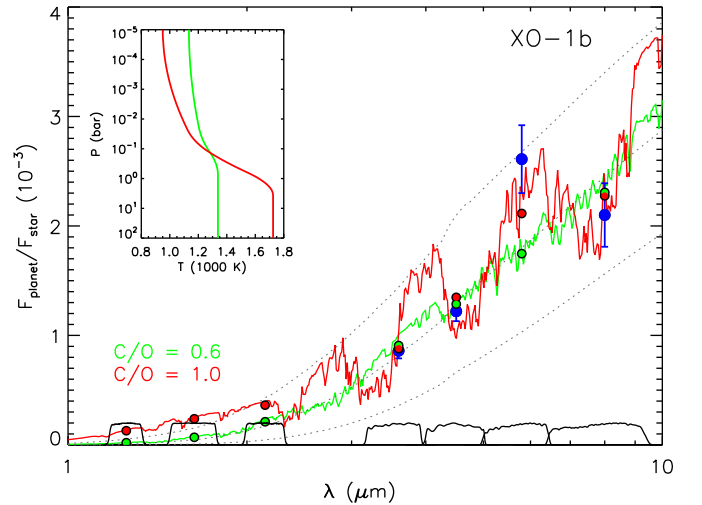}
\caption{Observations and models of thermal emission from the dayside atmosphere of XO-1b. The blue circles with error bars in the main panel show {\it Spitzer} IRAC observations from Machalek et al. (2008), and the solid curves show  two model spectra. The green curve shows a model with oxygen-rich (C/O = 0.5) composition, and the red curves shows 
a carbon-rich model with C/O $=$ 1. The carbon-rich model provides a better fit to the data. The oxygen-rich model under-predicts the flux at 5.8 $\micron$. The bandpass integrated model points are shown in the colored circles. The corresponding pressure-temperature ($P$-$T$) profiles, which do not have thermal inversions, are shown in the inset. The black curves at the bottom of the figure show seven photometric bandpasses: $J$, $H$, $Ks$, and {\it Spitzer} IRAC bands at 3.6 $\micron$, 4.5 $\micron$, 5.8 $\micron$, and 8 $\micron$. The dotted lines show three blackbody spectra of the planet at 1000 K, 1250 K, and 1500 K.} 
\label{fig:xo1b}
\end{figure}

\subsection{XO-1b}

The hot Jupiter XO-1b  (McCullough et al. 2006) has presented one of the extreme counterexamples to 
the TiO/VO hypothesis to date. The planet orbits a G dwarf on a 4 day orbit, and is one of the least irradiated 
hot Jupiters whose atmospheres have been observed to date, with an equilibrium temperature of $\sim$ 1400 K. 
As such, Fortney et al. (2008) predicted that the dayside atmosphere of XO-1b would not be able to host a thermal 
inversion since it would not be hot enough to maintain gaseous TiO or VO in the upper atmosphere. Subsequently, 
Machalek et al. (2008) reported observations of thermal emission from XO-1b in the four channels of {\it Spitzer} IRAC photometry (at 3.6$\micron$, 4.5$\micron$, 5.8$\micron$, and 8$\micron$). Interpreting the data using models 
of Burrows et al. (2008), Machalek et al. (2008) reported a thermal inversion in the dayside atmosphere of XO-1b. 
Given the low irradiation received at its substellar point, the inference of a thermal inversion in XO-1b presented a 
stark violation of the TiO/VO hypothesis. 

One potential explanation to the apparent contradiction was proposed by Knutson et al. (2010) who found a correlation 
between inferences of thermal inversions in hot Jupiters and the chomospheric activities of their host stars. They concluded that planets orbiting chromospherically active stars did not show evidence for thermal inversions in their atmospheres, and those orbiting quiet stars showed evidence for thermal inversions. Given the low chromospheric activity of its host star, the thermal inversion in XO1b appeared to be consistent with the empirical hypothesis of 
Knutson et al. However, it is still unknown to date as to what molecules in the atmosphere could actually cause 
a thermal inversion in XO-1b that is strong enough to match the observations of Machalek et al. (2008), since gaseous 
TiO and VO would be clearly underabundant in such a cool atmosphere (Spiegel et al. 2009). In order to create the 
thermal inversion in their models, Machalek et al. had to use an artificial opacity source in visible wavelengths with a 
parametric opacity, wavelength range, and location in the atmosphere (Burrows et al. 2008). Whether such a source is physically plausible in XO-1b is yet to be demonstrated. 

In the present work, we report a consistent explanation to the observations of Machalek et al. (2008). The observations and two model spectra are shown in Fig.~\ref{fig:xo1b} and the corresponding chemical compositions are shown in Table~\ref{tab:chem}. The four IRAC observations are shown in blue. Firstly, we confirm the findings of Machalek et al. that a solar abundance model cannot fit the data without a thermal inversion, as shown in the green curve in Fig.~\ref{fig:xo1b}. The model fits the data at 3.6 $\micron$,  4.5 $\micron$, and 8 $\micron$ to within the 1-$\sigma$ error bars, but under-predicts the 5.8 $\micron$ flux by $\sim$ 3-$\sigma$. Our non-inverted solar abundance model provides a slightly better fit to the data than that of Machalek et al. (2008), while still grossly under-predicting the flux at 5.8 $\micron$, in agreement with Machalek et al. Furthermore, the temperature structure required for a best-fit non-inverted solar abundance model as shown in Fig.~\ref{fig:xo1b} is nearly an isotherm, implying that the spectrum is equivalent to a blackbody, as shown by the dotted lines, making the O-rich composition irrelevant. The requirement of a thermal inversion with a solar abundance model thus rests largely on the flux observed in the 5.8 $\micron$ IRAC bandpass which has strong spectral features of H$_2$O. In a solar composition model, H$_2$O is the dominant oxygen bearing species at the temperatures of XO-1b. If there is no thermal inversion in the atmosphere, in which case temperature monotonically decreases outwards, a low flux is observed in the 5.8 $\micron$ IRAC band due to strong water absorption. The contrary observation of the high  5.8 $\micron$ flux in the data implies, in the context of a solar abundance model, that it must be due H$_2$O in emission rather than absorption. This is possible in the presence of a thermal inversion\footnote{A detailed description of such inferences is presented in Madhusudhan \& Seager, 2010).}. 

We report, however, that the observations of XO-1b can be explained without the requirement of an ad hoc thermal inversion if the atmosphere of XO-1b is carbon-rich (C/O $\geq 1$). A model spectra of XO-1b with a C-rich composition and a non-inverted temperature profile, is shown in red in Fig.~\ref{fig:xo1b}. The model fits all the observations to within the 1$\sigma$ uncertainties, on average. The key to explaining the high flux in the 5.8$\micron$ channel, simultaneously with the other photometric points, without a thermal inversion lies in the H$_2$O and CH$_4$ abundances expected in a C-rich atmosphere (see e.g. Madhusudhan et al. 2011). At the temperatures of XO-1b, C/O $\ge 1$ causes $\sim10-100$ times depletion of H$_2$O and $\sim 10-100$ times enhancement of CH$_4$ compared to a solar abundance model (with C/O = 0.5). Additionally, other hydrocarbons such as HCN and C$_2$H$_2$ are also expected to be abundant. The low H$_2$O abundance causes low absorption in the 5.8$\micron$ channel, leading to a high thermal flux emitted from the deeper layers of the atmosphere. On the other hand, the low fluxes in the 3.6 $\micron$ and 8 $\micron$ channels are explained by enhanced CH$_4$, HCN, and C$_2$H$_2$ absorption in both channels. The 4.5 $\micron$ flux in the C-rich model is explained by absorption due to CO and HCN. Therefore, we classify XO-1b as a C2 class planet in the framework of the 2-D classification scheme presented in section~\ref{sec:2d}. 

Future observations in the near-infrared with existing facilities will be required to confirm the C2 classification of XO-1b. 
Given that our current inference is based solely on the one photometric point at 5.8 $\micron$, new observations are required to confirm the potentially C-rich nature of XO-1b. As shown in Fig.~\ref{fig:xo1b}, the high C/O models fitting 
the IRAC data predict substantially higher fluxes in the 2-3 $\micron$ range compared to a solar composition 
model. This difference is largely due to different continuum emission between the two models; the high C/O 
solutions have higher temperatures in the low atmosphere ($P \gtrsim 1$ bar) which are required to fit the 
high 5.8 $\micron$ IRAC point. Such differences can be used to constrain the models using ground-based high precision photometry (e.g. Lopez-Morales, 2010; Croll et al. 2011) and spectrophotometry (Bean et al. 2010; Mandell et al. 2011; Crossfield et al. 2011), and space-borne spectroscopy with {\it HST} in the near term, and with {\it JWST} in the long term. 
Note that precisions better that $10^{-4}$ would be required to distinquish between the C-rich and O-rich models. 

Care must be taken, however, in the interpretation of the C/O of XO-1b using transmission spectroscopy. Given the low irradiation levels received by the planet, a transmission spectrum of XO-1b is likely to probe even cooler regions of the atmosphere.  It is possible that the temperatures in such regions can be well below the 1200 K minimum required for high-confidence estimates of C/O ratios (see Section~3.1). Consequently, spectra of the limb of the atmosphere may show absorption features of H$_2$O even if the C/O $\geq$ 1. In such cases, establishing the C/O ratio accurately 
would require estimating the H$_2$O and CH$_4$ abundances to precisions within factors of a few. Such precisions would require extremely precise (~10$^{-5}$) spectroscopic observations which may not be feasible with current instruments. At present, one could attempt to derive such molecular abundances using the {\it HST} NICMOS 
spectrum of XO-1b reported by Tinetti et al. (2010), whose reported constraints agree with a potential C-rich 
atmosphere.  However, we refrain from interpreting their data in the present work since their data reduction 
is currently contested (Gibson et al. 2010). 

\begin{figure}
\centering
\includegraphics[width = 0.5\textwidth]{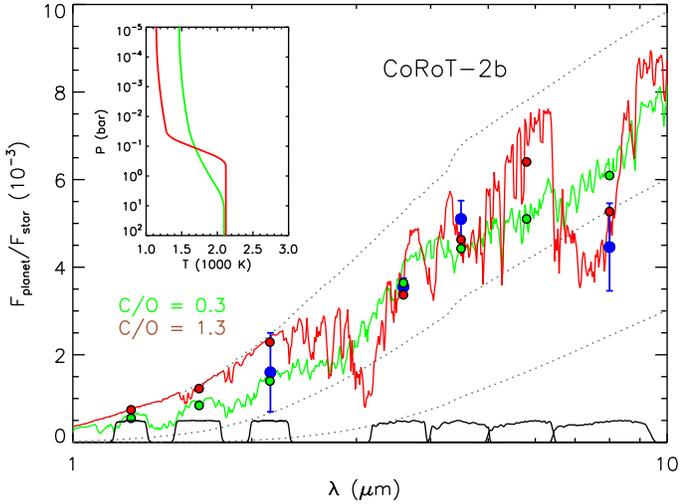}
\caption{Observations and model spectra of thermal emission for CoRoT-2b. The four observations are shown in blue circles with error bars at wavelengths 2.15 $\micron$ (Alonso et al. 2010), 3.6 $\micron$ (Deming et al. 2011), 4.5 $\micron$, and  8 $\micron$ (Gillon et al. 2010 \& Deming et al. 2011). The green curve in the main panel shows a model with an oxygen-rich (C/O < 1) atmosphere, and the red curve shows a model with a carbon-rich composition (C/O $\geq$  1). The corresponding pressure-temperature profiles are shown in the inset.  The temperature profiles do not have thermal inversions. The red and green colored circles show the corresponding models integrated in different photometric bandpasses, the filter functions for which are shown in black at the bottom. The dotted curves show three blackbody spectra of the planet at 1000 K, 1500 K, and 2100 K, for reference.} 
\label{fig:corot2b}
\end{figure}

\subsection{CoRoT-2b}

The hot Jupiter CoRoT-2b (Alonso et al. 2008) presents yet another challenge for atmospheric characterization of hot Jupiters using traditional models and classification schemes. The planet orbits a late G dwarf on a 1.7 day orbit and 
is one of the highly irradiated hot Jupiters predicted by to host a thermal inversion in its dayside atmosphere 
(Fortney et al. 2008). However, the host star is known to be very chromospherically active, which according to the 
Knutson et al (2010) hypothesis would suggest that there should not be a thermal inversion in its dayside atmosphere. 
Between the two competing predictions, however, lies the uncomfortable fact that neither models with thermal inversions nor models without thermal inversions have been able to explain the observed spectrum of CoRoT-2b (Gillon et al. 2010, Deming et al. 2011). Observations of dayside thermal emission from CoRoT-2b showed an anomalously low flux in the 8$\micron$ {\it Spitzer} IRAC channel, and high flux in the 4.5 $\micron$ channel, both of which could not be explained by previous atmospheric models (Alonso et al. 2010, Gillon et al. 2010, Deming et al. 2011, Guillot et al. 2011). As an alternative explanation, Deming et al. (2011) required either excess emission from tidal mass loss that is too high to be realistic or the presence of an unknown absorber that selectively absorbs at wavelengths below 5 $\micron$.

In the present work, we are able to explain all the existing observations of CoRoT-2b. Figure~\ref{fig:corot2b} 
shows the observations and two model spectra fitting the data, and the corresponding chemical compositions are shown in Table~\ref{tab:chem}. The data include broadband photometric 
detections of thermal emission from CoRoT-2b in four infrared channels: a ground-based Ks band at 2.1 $\micron$ (Alonso et al. 2010), and three {\it Spitzer} IRAC bandpasses at 3.6 $\micron$ (Deming et al. 2011), 4.5 $\micron$ and 8 $\micron$ (Gillon et al. 2010; Deming et al. 2011). The models shown in  Figure~\ref{fig:corot2b} include an O-rich model 
(C/O $<$ 1) and a carbon-rich (C/O $>$ 1). The corresponding temperature profiles are shown in the inset of Figure~\ref{fig:corot2b}. As shown in the $P$-$T$ profiles, the observations rule out a thermal inversion in the dayside 
atmosphere of CoRoT-2b. The low flux in the 8 $\micron$ IRAC channel, thought to be anomalous by previous 
studies, can be naturally explained by CH$_4$, HCN, and C$_2$H$_2$ absorption in the carbon-rich model 
and by H$_2$O absorption in the oxygen-rich model. The same opacity sources also explain the low flux in the 3.6 $\micron$ channel. 

Existing observations of CoRoT-2b are less constraining of its C/O ratio, implying that it could be classified as O1 or C2 according to the 2D classification scheme proposed in section~\ref{sec:2d}, although the C2 class is marginally favoured. 
We find that all the observations can be explained to within the 1-$\sigma$ uncertainties on average by a wide range of models with carbon-rich (C/O $>$ 1) and oxygen-rich (C/O $<$ 1) compositions. The {\it Spitzer} IRAC 5.8 $\micron$ flux that implied a substantial depletion of H$_2$O in XO-1b is unavailable for CoRoT-2b, leading to a poorer constraint on the H$_2$O abundance in the latter. Thus, it is expected that both H$_2$O-rich and H$_2$O-poor solutions are possible for CoRoT-2b, implying low C/O ($<1$) and high C/O ($>1$) ratios, respectively. This degeneracy occurs because H$_2$O, CH$_4$, and HCN, all have absorption features in the 3.6 and 8 $\micron$ IRAC channels. A 5.8 $\micron$ observation could have broken the degeneracy with an independent constraint on the H$_2$O abundance; the IRAC 5.8 $\micron$ bandpass does not contain significant CH$_4$ features. However, since the C-rich model provides a marginally better fit to the data, especially at 8 $\micron$, we tentatively classify CoRoT-2b as a C2 class planet in the framework of our 2D classification scheme proposed in section~\ref{sec:2d}. 

New observations of CoRoT-2b would be required to conclusively constrain its C/O ratio. As shown in Figure~\ref{fig:corot2b}, the two spectral models differ substantially in their absorption features. In particular, observations using the {\it HST} WFC3 instrument in the 1.1-1.7 $\micron$ bandpass would be able to conclusively determine the H$_2$O abundance based on the strong H$_2$O feature in the O-rich model around 1.4 $\micron$. In addition, ground-based observations with higher precision that that currently available in the Ks band would be able to better constrain the lower atmosphere temperature profile of the atmosphere, which in turn could differentiate between the different models. Another 
alternative would be to use transmission spectroscopy and photometry from ground and space to constrain the atmospheric composition of the day-night terminator of the planet. 

\begin{figure}
\centering
\includegraphics[width = 0.5\textwidth]{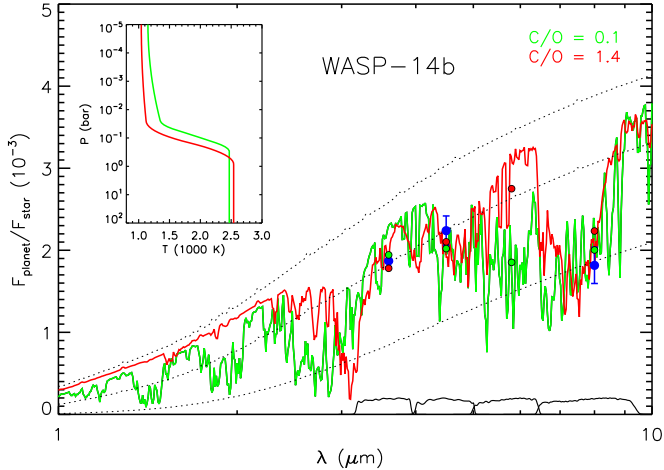}
\caption{Observations and model spectra of thermal emission from the dayside atmosphere of WASP-14b. The data are shown in blue circles with error bars obtained in the {\it Spitzer} IRAC bandpasses at 3.6 $\micron$, 4.5 $\micron$, and 8 $\micron$ (Blecic et al. 2012). The solid green and red curves show two model spectra with different C/O ratios. Both models provide reasonable fits to the data, though the oxygen-rich model provides a marginally better fit. The temperature profiles corresponding to the two models are shown in the inset. A thermal inversion is ruled out by the data. The {\it Spitzer} bandpasses are shown in the solid black curves at the bottom of the plot. The dotted gray curves show the planet blackbody spectra at 1600 K, 2200 K, and 2600K.} 
\label{fig:wasp14b}
\end{figure}

\subsection{WASP-14b}
\label{sec:wasp14b}

WASP-14b is a transiting hot Jupiter orbiting an F5 main-sequence star at an orbital separation 
of 0.036 AU (Joshi et al. 2009). With an incident irradiation of 2.5 $\times$ $10^9$ ergs/cm$^2$/s, leading to 
a zero-albedo equilibrium temperature of $\sim$ 2200K, WASP-14b belongs to the class of highly 
irradiated hot Jupiters that are expected to host strong thermal inversions based on the TiO/VO 
hypothesis of Fortney et al. (2008). However, using three channels of {\it Spitzer} IRAC photometry, at 
3.6 $\micron$, 4.5 $\micron$, and 5.8 $\micron$, Blecic et al. (2012) reported the lack of a thermal 
inversion in the dayside atmosphere of WASP-14b. The absence of a thermal inversion in WASP-14b 
was evident primarily from a low brightness temperature of $\sim$ 1600K in the 8 $\micron$ channel 
compared to temperatures above 2200 K observed in the lower wavelength channels at 
3.6 $\micron$ and 4.5 $\micron$, indicating temperature decreasing outward in the atmosphere. 

While the lack of a thermal inversion in WASP-14b may contradict the prediction of 
Fortney et al. (2008), it is also possible that condensation and gravitational settling of TiO/VO 
(Spiegel et al. 2009) might preclude the formation of a thermal inversion. It is unclear if photospheric 
activity of the host star may have caused the disappearance of TiO and VO. Knutson et al. (2010) find that 
host star WASP-14 is not very active implying that the planet should have had a thermal inversion. 
But Knutson et al. (2010) also highlight the difficulty in accurately establishing the Ca II H and K 
indices for F stars, such as WASP-14b. Alternately, as shown in Madhusudhan et al. (2011b) and in the 
present work, it may also be possible that the atmosphere of WASP-14b is carbon-rich (i.e. C/O $\geq$ 1) 
in which case TiO and VO are naturally low in abundance. 

We confirm the conclusions of Blecic et al. (2012) that the limited data available cannot conclusively 
constrain the C/O ratio of the dayside atmosphere of WASP-14b. Two model spectra fitting the observations 
are shown in Fig.~\ref{fig:wasp14b}, and Table~\ref{tab:chem} shows the model compositions. In the O-rich model, the 
dominant absorption in the 8-$\micron$ and 3.6-$\micron$ IRAC channels is caused by H$_2$O, whereas 
that in the C-rich model is caused by HCN, CH$_4$, and C$_2$H$_2$. Both models contain absorption 
features due to CO in the 4.5 $\micron$ bandpass. The temperature profiles for the two models are shown 
in the inset in Fig.~\ref{fig:wasp14b}. Since the data rule out a thermal inversion in the atmosphere 
regardless of the C/O ratio, both models have similar lower atmospheric temperatures. 

Given that both C-rich and O-rich models explain the data almost equally well, WASP-14b could be 
classified either as an O1 or a C2 system. Note that despite the high irradiation, WASP-14b cannot be 
classified as an O2 system because of the lack of a strong thermal inversion and hence falls in the 
gray area between the O1 and O2 classes shown in Fig.~\ref{fig:2d}. However, since an oxygen-rich 
model (in green) provides a marginally better fit to the 8-$\micron$ IRAC point than the carbon-rich 
model (in red), one may tentatively classify WASP-14b as an upper-O1 class planet, as shown in Fig.~\ref{fig:2d_1}. 

New observations are needed to establish the C/O ratio of WASP-14b. The degeneracy 
between the two models arises from the lack of a spectral measurement in the IRAC 5.8 $\micron$ channel. 
In principle, a 5.8-$\micron$ IRAC data point, if available, could have easily removed the 
degeneracy. Given that the 5.8-$\micron$ IRAC channel is no longer available on {\it Spitzer}, an 
alternative way to constrain the water abundance with existing facilities would be to observe 
the H$_2$O bands in the {\it HST} WFC3 bandpass between 1.1 and 1.7 $\micron$.

Future confirmation of the C/O ratio of WASP-14b would lead to one of two interesting conclusions. 
If the C/O is found to be less than unity, based on {\it HST} observations, it will be a stringent constraint 
on the irradiation evidence for gravitational settling of TiO/VO according to the the Spiegel et al (2009), 
and will place a lower-limit on the irradiation level that defines the gray region between the O1 and O2 
classes in Figs.~\ref{fig:2d} \& \ref{fig:2d_1}. 
Alternately, if the C/O is found to be $\geq$ 1, WASP-14b will join WASP-12b in the new class of 
CRPs or carbon-rich planets (Madhusudhan et al. 2011b) . 

\begin{figure}
\centering
\includegraphics[width = 0.5\textwidth]{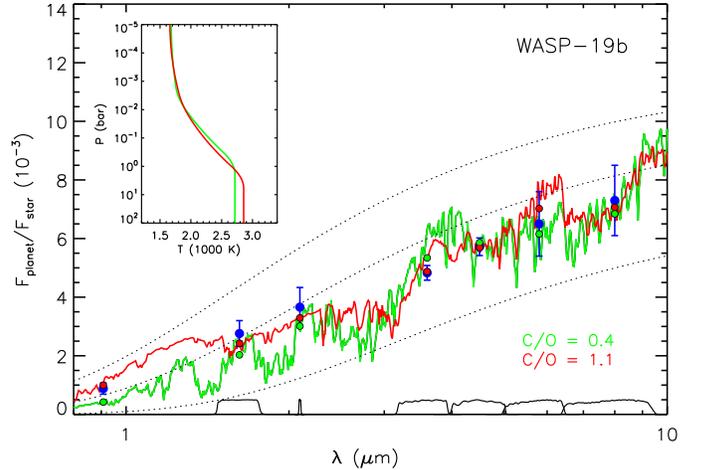}
\caption{Observations and model spectra of the dayside atmosphere of WASP-19b. The observations, shown in blue circles with error bars, comprise of three ground-based photometric data at 0.9 $\micron$ (Burton et al. 2012), 
1.6 $\micron$ (Anderson et al. 2010) and 2.1 $\micron$ (Gibson et al. 2010), and four photometric observations in the 
{\it Spitzer} IRAC bandpasses at 3.6 $\micron$, 4.5 $\micron$, 5.8 $\micron$, and 8 $\micron$  (Anderson et al. 2012). 
The solid red and green curves in the main panel show two model spectra with different C/O ratios as described in the legend. The data marginally favor a carbon-rich model (C/O $\geq$ 1), though an oxygen-rich model (C/O $<$ 1)  cannot be conclusively ruled out. The corresponding temperature profiles are shown in the inset. A strong thermal inversion is ruled out by the data. The dotted lines show three blackbody spectra at 1800 K, 2500 K, and 2900 K. The black curves at the bottom of the figure show the photometric bandpasses corresponding to the data.} 
\label{fig:wasp19b} 
\end{figure}

\subsection{WASP-19b}
\label{sec:wasp19b}

The transiting hot Jupiter WASP-19b orbits a G8 main-sequence star at a separation of 0.017 AU (Hebb et al. 2010). 
At an irradiation level of 4 $\times$ $10^9$ ergs/cm$^2$/s, it is one of the very highly irradiated hot Jupiters with 
a zero-albedo equilibrium temperature of 2400 K. Consequently, it is also expected to host a strong thermal 
inversion due to TiO/VO if the atmosphere is in chemical equilibrium and the composition is oxygen-rich as 
predicted by Fortney et al. (2008). However, using six channels of broadband photometry obtained using 
{\it Spitzer} and ground-based facilities, Anderson et al. (2012) report the non-detection of a thermal inversion 
in the dayside atmosphere of WASP-19b. Their inference was based on observations of thermal emission in 
four channels of {\it Spitzer} IRAC photometry at 3.6 $\micron$, 4.5 $\micron$, 5.8 $\micron$, and 8 $\micron$, and 
two ground-based photometric measurements at 1.6 $\micron$ (Anderson et al. 2010) and 2.1 $\micron$ (Gibson et al. 2010). The inference of a non-inverted profile is evident from the higher brightness temperatures observed in 
the 1.6-3.6 $\micron$ bands compared to the longer wavelength bands, suggesting a temperature decreasing 
outward in the atmosphere. 

Even though the lack of a noticeable thermal inversion in WASP-19b defies the predictions of the 
TiO/VO hypothesis of Fortney et al. (2008), alternate explanations exist. It might be unlikely that gravitational 
settling (Spiegel et al. 2009) alone can deplete all the TiO and VO in this highly irradiated atmosphere. 
However, Anderson et al. (2012) note that the planet orbits a chromospherically active star, in which case, 
according to the hypothesis of Knutson et al. (2010) the high UV flux incident on the planetary 
atmosphere could potentially dissociate the inversion causing molecules. An alternative explanation, as suggested by Madhusudhan et al. (2011), could be that 
the atmosphere is carbon-rich in which case TiO and VO would be naturally under-abundant and hence the 
lack of a thermal inversion. Therefore, WASP-19b could belong to either class O1/O2 or C2, depending on the C/O ratio. 

Current observations of WASP-19b are inconclusive about its C/O ratio but nominally favor a C/O $\geq$ 1. The 
data and two model spectra of WASP-19b are shown in Fig.~\ref{fig:wasp19b} (see Table~\ref{tab:chem} for the chemical compositions of the models). As demonstrated by Anderson et al. (2012), the majority of available data are fit equally well by O-rich as well as C-rich models. The large error bar on the flux in the 5.8 $\micron$ {\it Spitzer} IRAC channel makes it difficult to constrain the H$_2$O abundance which is one of the 
major discriminator between the O-rich and C-rich models. And, as in the case of WASP-14b, the remaining three {\it Spitzer} data at 3.6 $\micron$, 4.5 $\micron$, and 5.8 $\micron$ are also fit equally well by both models, although the 3.6 $\micron$ flux is better explained by the C-rich model. The opacity sources in the different IRAC channels are the same as those discussed for WASP-14b. The photometric data at 1.6 $\micron$ and 2.1 $\micron$ do not particularly discriminate between the two models because both models have similar lower atmospheric temperatures, since neither of them hosts a thermal inversion. 

However, an important discrimination is provided by the z$^\prime$-band observation (Burton et al. 2012) centered 
at 0.9 $\micron$, in which TiO is a prominent opacity source for the O-rich model. Even though TiO and VO may not be abundant enough in the upper-atmosphere, due to either gravitational settling or chromospheric activity, in an O-rich atmosphere it should still be present in the lower atmosphere around $P \sim 1$ bar where the temperatures are hottest and where the incident irradiation doesn't reach with significant intensity. Consequently, if C/O < 1 and there is no thermal inversion, highly irradiated atmospheres like WASP-19b should display noticeable TiO and VO absorption between 0.7 and $\sim1.1 micron$, which is observable as decreased flux in the z$^\prime$, $z$, and $J$ bands, as can be seen in Fig.~\ref{fig:wasp19b}.  The flux observed in the z$^\prime$ band is better fit by the C-rich model which doesn't have significant TiO/VO absorption. Consequently, we tentatively classify WASP-19b as a C2 class planet in Fig.~\ref{fig:2d_1} in view of the marginally better fit provided by the C-rich model at 0.9 $\micron$ and 3.6 $\micron$.

New observations are required to conclusively determine the C/O ratio of WASP-19b and to classify it 
as O2 or C2. As shown in Fig.~\ref{fig:wasp19b},  new ground-based data in the $z$ and $J$ bands will be able to 
discriminate between the different flux continua at 1 $\micron$ and 1.2 $\micron$ predicted by the O-rich and C-rich models. Additionally, observations with {\it HST} WFC3 would be able to provide a robust discrimination between the 
two models in the H$_2$O bands between 1.1 $\micron$ and 1.7 $\micron$. 

\begin{figure}
\centering
\includegraphics[width = 0.5\textwidth]{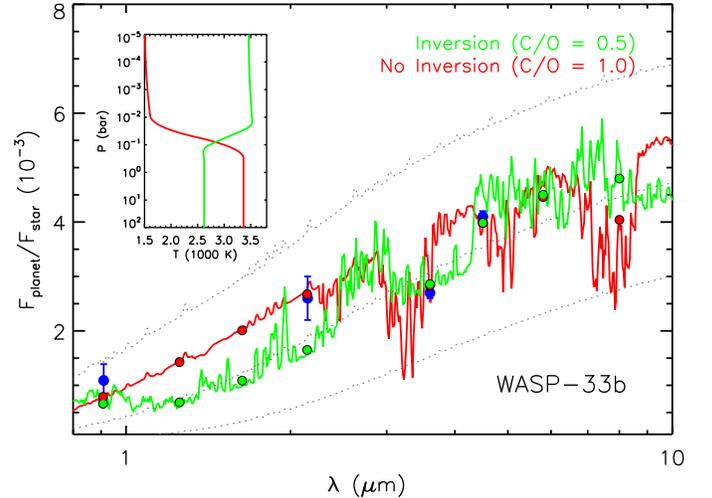}
\caption{Observations and model spectra of the dayside atmosphere of hot Jupiter WASP-33b. The observed planet-star flux contrasts are shown in blue and include photometric data from ground-based observations at 0.9 $\micron$ (Smith et al. 2011) and at 2.1 $\micron$ (Deming et al. 2012), and space-based {\it Spitzer} observations in the IRAC bandpasses at 3.6 $\micron$ and 4.5 $\micron$ (Deming et al. 2012). Two model spectra with different C/O ratios and temperature profiles are shown in the solid curves. The data are best explained by a carbon-rich model without a thermal inversion, shown in red. However, an oxygen-rich model with a thermal inversion can also explain all but the 2.1 $\micron$ point. The dotted grey curves show blackbody spectra of the planet at 2000 K, 2800 K, and 3800 K.} 
\label{fig:wasp33b}
\end{figure}

\subsection{WASP-33b} 
\label{sec:wasp33b}

The transiting hot Jupiter WASP-33b, orbiting an A5 main-sequence star at an orbital separation of 0.03 AU is one of 
the most highly irradiated hot Jupiters known ( Collier-Cameron et al. 2010; Herero et al. 2011). At an incident 
irradiation exceeding $10^{10}$ ergs/cm$^2$/s, and an equilibrium temperature exceeding 3300 K, it the hottest planet in our sample and is comparable in temperature to low mass stars. Consequently, WASP-33b is the most suitable hot Jupiter to study thermal and chemical characteristics of irradiated giant planets. If the atmosphere of 
WASP-33b is oxygen-rich, TiO and VO should be abundant in the atmosphere, despite any gravitational settling. 
Consequently, an oxygen-rich WASP-33b would offer the most unambiguous opportunity to detect a thermal inversion in a hot Jupiter atmosphere and place it in the O2 class discussed in section~\ref{sec:2d}. On the other hand, the lack of a thermal inversion in WASP-33b would be a strong indication of a carbon-rich atmosphere, i.e a C2-class planet. 

Observations of the dayside atmosphere of WASP-33b nominally allow both possibilities, O2 as well as C2, 
thus necessitating new observations to break the degeneracy. The available data comprise of ground-based 
photometry in the z$^\prime$ band (Smith et al. 2012) and the K$_s$ band (Deming et al. 2012), and 
{\it Spitzer} IRAC photometry in the 3.6 $\micron$ and 4.5 $\micron$ channels (Deming et al. 2012). 
Deming et al. (2012) conclude that all the existing observations are consistent with either a carbon-rich 
atmosphere without a thermal inversion (i.e. a C2-type atmosphere) or an oxygen-rich atmosphere with 
a thermal inversion (i.e. an O2-type atmosphere). While our results in the present work generally agree 
with the assessment of Deming et al., we note that a C2-type atmosphere for WASP-33b explains the data 
marginally better than one of O2 type. The data and two model spectra of WASP-33b are shown in 
Fig.~\ref{fig:wasp33b}. Table~\ref{tab:chem} shows the chemical compositions of the models. 

The data in the z-band and in the two {\it Spitzer} bands, at 3.6 $\micron$ and 4.5 $\micron$, can be explained 
almost equally well by both models (C2 and O2), but for very different reasons. The high flux observed in the 
 z$^\prime$ band is naturally explained by the higher lower atmospheric temperature in the non-inversion model. 
 And since that model is C-rich, no significant TiO or VO absorption is expected and hence the high predicted 
 flux in the z$^\prime$ band. On the other hand, the O2 model which has a thermal inversion has a cooler lower 
 lower atmosphere than the non-inverted model. However, because this model is O-rich substantial TiO and VO is 
 present and is seen as an emission feature, as opposed to absorption, due to the thermal inversion. Consequently, 
 the O2 model predicts a higher brightness temperature in the z$^\prime$ band than its lower atmospheric temperature 
 and hence explains the high observed z$^\prime$ band flux. Similar reasoning explains the fits of both models to the {\it Spitzer} data at 3.6 $\micron$ and 4.5 $\micron$. At 3.6 $\micron$ the low observed flux is explained by the inversion model based on the lower temperature of its lower atmosphere and lack of any significant opacity source at 3.6 $\micron$ in its oxygen-rich atmosphere for $T \geq $ 2500 K. The same model explains the higher 4.5 $\micron$ flux due to CO in emission. On the other hand, the non-inverted C-rich model explains the 3.6 $\micron$ point due to strong absorption by HCN, CH$_4$ and C$_2$H$_2$ in the 3.6 $\micron$ channel and absorption due to CO and HCN in the 4.5 $\micron$ channel. 

The K$_s$-band flux, however,  favors a C-rich atmosphere without a thermal inversion. 
The O2 model predicts over 2-$\sigma$ lower planet-star flux contrast than that observed, whereas a 
C2 model fits the data point rather precisely, within the 1-$\sigma$ errors. The K$_s$ band is devoid 
of strong molecular absorption and hence probes thermal emission from the deepest layer of a planetary 
atmosphere ($P \sim 1$ bar) until continuum opacity makes the atmosphere optically thick. Consequently, 
the difference between the K$_s$ band fluxes predicted by the two models directly reflects the difference 
between the isothermal temperatures of the deep atmosphere in the two models. More generally, since 
temperature profiles with thermal inversions have cooler lower atmospheres compared to those without 
thermal inversions, for the same irradiation, thermal emission in opacity windows such as the $J$, $H$, 
and $K$ bands tends to be typically lower for models with thermal inversion. 

WASP-33b is one of the most ideal targets for follow-up observations with existing instruments.  
The very high planet-star flux contrast implies that thermal emission from the planet can easily 
be detected with routine observations in the near-IR as has already been demonstrated by Smith et al  and 
Deming et al. New ground-based observations in the J-band (1.2 $\micron$) and H-band (1.6 $\micron$) 
and {\it HST} observations with the WFC3 instrument should be able to conclusively distinguish between 
the two possibilities of an O2 and a C2 atmosphere. As shown in Fig.~\ref{fig:wasp33b}, $J$ and $H$ band 
observations should be able to place tight constraints on the temperature in the lower atmosphere and hence 
the presence of a thermal inversion, given the large difference in flux contrast between the two models in 
those bands. Furthermore, an oxygen-rich WASP-33b must show strong emission features of H$_2$O 
in the {\it HST} WFC3 bandpass due to a thermal inversion in accordance with the TiO/VO hypothesis 
of Fortney et al. (2008). On the other hand, weak features in the water bands would indicate a 
carbon-rich atmosphere in WASP-33b.

\begin{figure}
\centering
\includegraphics[width = 0.5\textwidth]{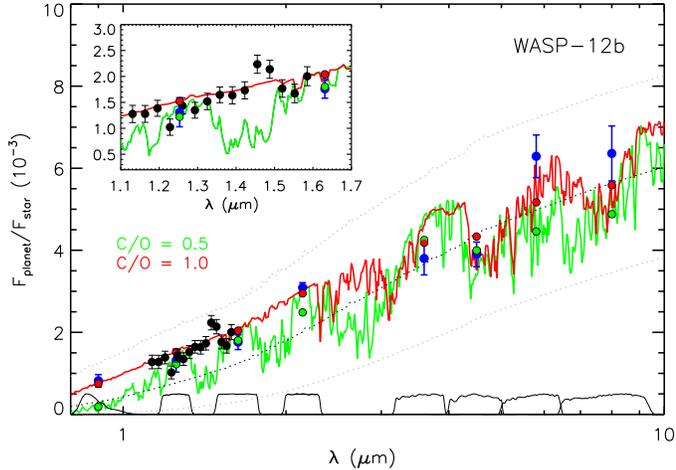}
\caption{Observations and model spectra of dayside thermal emission from the hot Jupiter WASP-12b. The data are shown in blue and black circles with error bars. The blue data show ground-based broadband photometric points at 0.9 $\micron$ (Lopez-Morales et al. 2010), 1.2 $\micron$, 1.6 $\micron$, and 2.1 $\micron$ (Croll et al. 2010), and {\it Spitzer} IRAC photometry at 3.6 $\micron$, 4.5 $\micron$, 5.8 $\micron$ and 8 $\micron$ (Campo et al. 2011, Cowan et al. 2012). The data in black circles show a near-infrared spectrum obtained using the {\it HST} Wide Field Camera 3 
(Swain et al. 2012). The photometric bandpasses are shown in black solid curves at the bottom of the plot and the {\it HST} WFC3 bandpass is shown in the inset. The solid curves show two models with different C/O ratios. The data are explained well by a carbon-rich model, whereas a model with oxygen-rich composition is ruled out by the data, consistent with the findings of Madhusudhan et al. (2011a). Both models have the same  temperature profile without a thermal inversion; a strong thermal inversion is ruled out by the data. The dotted gray curves show the planet blackbody spectra at 1800 K, 2500 K, and 3200 K.} 
\label{fig:wasp12b}
\end{figure}

\subsection{WASP-12b}
\label{sec:wasp12b}

The atmosphere of the transiting hot Jupiter WASP-12b (Hebb et al. 2009) has generated substantial 
interest in the recent past. Being one of the most highly irradiated hot Jupiters, with an equilibrium 
temperature of $\sim$ 2600 K, initial models assuming solar abundance (i.e. O-rich) compositions predicted 
strong thermal inversions to be present in the dayside atmospheres of planets like WASP-12b 
(Fortney et al. 2008). Even considering depletion due to gravitational settling, TiO was expected 
to be abundant in the dayside atmosphere of WASP-12b and, hence, cause a thermal inversion (Spiegel et al. 2009). 
Furthermore, an O-rich atmosphere would imply the presence of abundant H$_2$O and negligible 
CH$_4$ (Burrows \& Sharp 1999). However, observations of the dayside atmosphere of WASP-12b 
suggested a stark contrast to previous theoretical predictions which assumed solar abundances. 
Using multi-band photometric observations over a wide wavelength range (1-10 $\micron$) obtained 
with {\it Spitzer} (Campo et al. 2011) and from ground (Croll et al. 2010), Madhusudhan et al. (2011a) 
reported the detection of a carbon-rich atmosphere (C/O $\geq$ 1), implied by an under-abundance 
of H$_2$O  and over-abundance of CH$_4$, and the lack of a strong thermal inversion in the dayside 
atmosphere of WASP-12b. Consequently, it is apparent that WASP-12b cannot be characterized solely 
on the basis of incident irradiation, and assuming a solar C/O ratio, as per previous classification schemes 
(e.g. Fortney et al. 2008). However, the C/O $\geq$ 1 and the lack of a strong thermal inversion in WASP-12b 
are both consistent with the properties of a C2-class planet according to our 2-D classification scheme 
suggested in section~\ref{sec:2d}. 

New data confirm a high C/O ratio and a weak thermal inversion in the dayside atmosphere of WASP-12b. Using new observations in the {\it Spitzer} IRAC channels at 3.6 $\micron$ and 4.5 $\micron$, in addition to existing data, Cowan et al. (2012) confirmed that their best fitting models required a high C/O ratio and the lack of a strong thermal inversion in the dayside atmosphere of WASP-12b (see Section~5.2.1 of Cowan et al.). It is to be noted, however, that since Cowan et al. used forward models (Burrows et al. 2008) for their interpretation, as opposed to a retrieval algorithm of Madhusudhan et al, a statistical estimate of the C/O ratio was not possible in the Cowan et al study, despite their confirmation of a C/O $\geq 1$ implied by their high CO/H$_2$O requirement. New studies have also reported ground-based photometric (Zhao et al. 2012) and spectroscopic (Crossfield) observations of thermal emission from WASP-12b in the K band (around 2.1 $\micron$), and confirmed previous measurements in the same bandpass by Croll et al. (2010).  More recently, Swain et al. (2012) reported observations of WASP-12b in the near-infrared (1-1.7 $\micron$) using the {\it HST} Wide Field Camera 3 (WFC3). Based on their thermal emission spectra, Swain et al. suggested that even though models with C/O $\geq$ 1 best explained their data, they could not provide conclusive constraints on the C/O ratio. However, Swain et al. stress that explaining their data required a substantial paucity of H$_2$O ($\lesssim 10^{-6}$). It is well known that such low levels of H$_2$O is possible if either C/O $\geq$ 1, or if the overall metallicity of the planet is low (e.g. see Section~3; also see Seager et al. 2005, Madhusudhan et al. 2011, Moses et al. 2012). In the latter case, both C/H and O/H would have to be under-abundant in WASP-12b by about two orders of magnitude below the stellar values, a scenario which would not be able to explain the high CO required to explain the {\it Spitzer} data in Madhusudhan et al. (2011a) and Cowan et al. (2012), which Swain et al did not consider including in their model fits. Furthermore, a hundred times lower metallicity (particularly C and O abundances) in a giant planetary atmosphere relative to its host star would be challenging to explain using the standard core accretion model for giant planet formation (see e.g. Owen et al. 1999; Atreya \& Wong 2005). 

In the present work, we find that all existing observations of thermal emission from WASP-12b are 
consistent with a high C/O ratio and a weak thermal inversion in its dayside atmosphere, confirming the 
findings of Madhusudhan et al. (2011a). The data include four channels of ground-based near-infrared 
photometry in the $z^\prime$ (Lopez-Morales et al. 2010), $J$, $H$, and $K$ bands (Croll et al. 2011), 
four-channels of {\it Spitzer} IRAC photometry (Campo et al. 2011; for 3.6 $\micron$ and 4.5 $\micron$ we use 
the new values from Cowan et al. 2012), and a near-infrared spectrum in the {\it HST} WFC3 bandpass (Swain et al. 2012). We have also updated our spectral models, over those used in Madhusudhan et al. (2011), by including 
molecular opacities due to C$_2$H$_2$ and HCN, which could be important sources of opacity in carbon-rich 
atmospheres (see Section~3). Fig.~\ref{fig:wasp12b} shows all the data along with two spectral 
models, and the chemical compositions of the models are shown in Table~\ref{tab:chem}. We find that a carbon-rich model (with C/O $\geq$ 1) provides the best fit to all the data whereas a solar abundance model is inconsistent with the data,  in agreement with the interpretation of Madhusudhan et al. (2011a). The primary difference between the fits in Madhusudhan et al. (2011a) and the present work lies in the interpretation of the absorption in the 8$\micron$ IRAC channel. While Madhusudhan et al. (2011a) suggested that the low flux at 8$\micron$ was solely due to CH$_4$ absorption, we find that the low flux could be due to a combination of CH$_4$ and HCN opacity. The uncertainties in high 
temperature opacities of hydrocarbons (Tinetti et al. 2010) makes it difficult to judge the exact relative 
contributions of CH$_4$ and HCN, but it is possible that in C-rich atmospheres HCN may 
dominate the 8$\micron$ opacity, as has also been suggested by Kopparapu et al. (2012) and Moses et al. (2012). 

We classify WASP-12b as a C2 class planet, based on all observations of its dayside atmosphere. The {\it HST} WFC3 spectrum reported by Swain et al. (2012) provides a particularly unambiguous constraint on the C/O ratio as shown in the inset of Fig.~\ref{fig:wasp12b}. The WFC3 data by themselves alone conclusively rule out H$_2$O absorption (as also reported by Swain et al. 2012) which is a prominent feature around 1.4 $\micron$, and hence rule out an O-rich atmosphere. Furthermore, the small absorption feature in the data at 1.57 $\micron$ appears to  be consistent with the small C$_2$H$_2$ feature in the C-rich model, although the detection is only marginal. The data also rule out a strong thermal inversion in the atmosphere, which is apparent from the lack of any strong emission features in the spectrum, as shown in Fig.~\ref{fig:wasp12b}. 

Additional constraints on the atmospheric composition of WASP-12b are expected in future work. In this work, we have not considered observations of transmission spectra of WASP-12b which probe the atmospheric composition at the day-night terminator of the planet. Cowan et al. (2012) reported transmission photometry using {\it Spitzer} at 3.6 $\micron$ and 4.5 $\micron$. They observed unusually high absorption in the 3.6 $\micron$ channel compared to the the 4.5 $\micron$ channel, which they were unable to explain using any composition in their atmospheric models, in which the dominant sources of opacity were only H$_2$O and CO, and, probably, CH$_4$. Swain et al. (2012) reported transmission spectroscopy of WASP-12b using {\it HST} WFC3. In order to explain their data, Swain et al. required  extremely low levels of H$_2$O ($\lesssim 5 \times 10^{-5}$) and a possible presence of TiH and CrH in the atmosphere. 
Such a composition is possible if either the atmospheric metallicity is at least two orders of magnitude lower than the stellar metallicity or if the metallicity is normal but the C/O ratio is $\geq 1$ as has been inferred on the dayside by Madhusudhan et al. (2011) and in the present work. Future observations of WASP-12b in transit in the near-infrared (between 0.7 $\micron$ and 1.1 $\micron$) could also place constraints on the TiO/VO abundances in its atmosphere.
In an upcoming effort, we plan to reanalyze all existing transmission data of WASP-12b and fit them with our models which include a broader set of opacities, than previous studies, including HCN and C$_2$H$_2$, to obtain statistical constraints on the atmospheric C/O ratio at the day-night terminator of WASP-12b. 

\begin{figure}[]
\centering
\includegraphics[width = 0.5\textwidth]{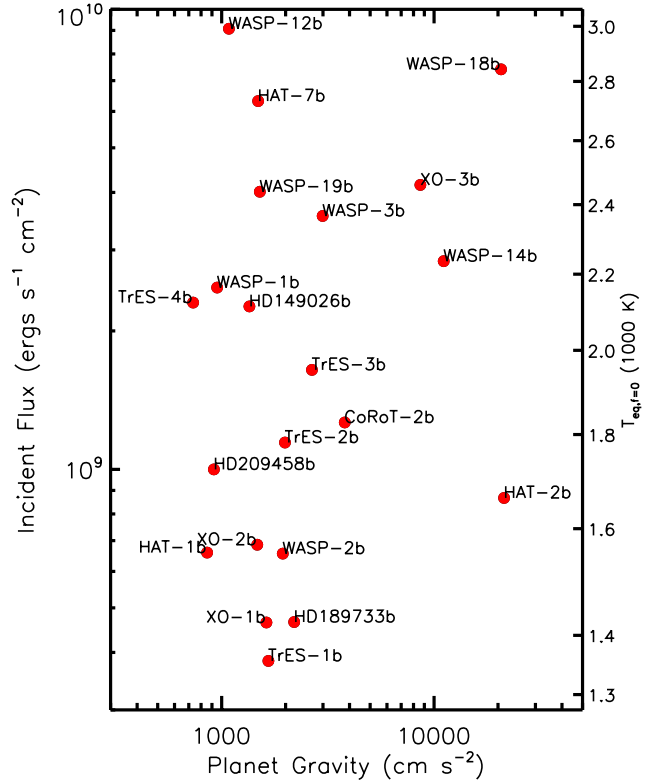}
\caption{Transiting hot Jupiters that are optimal for characterizing atmospheric C/O ratios (see Section~\ref{sec:obs_planning}). The left vertical axis shows the incident flux from the host star received by the planet at the sub-stellar point. The right vertical axis shows the equilibrium temperature with zero albedo and no redistribution ($T_{\rm eq,f=0}$), which is a nominal lower limit on the quench temperature ($T_q$), as discussed in Section~\ref{sec:quench}. All the 21 systems shown here have temperatures greater than 1300 K in their lower atmospheres, at $P \sim 1$ bar, and have all been observed in at least three channels of {\it Spitzer} photometry. Combining existing data of these systems with new observations in the near-infrared, that are possible with existing facilities, can yield 
constraints on their C/O ratios.The systems parameters for the planets are adopted from http://www.exoplanets.org. } 
\label{fig:irac_systems}
\end{figure}

\section{Observing Strategy and Optimal Systems}
\label{sec:obs_planning}

Future observations in the near-infrared with existing facilities will be able to constrain C/O ratios for several 
currently known exoplanets. As discussed for each system in Section~\ref{sec:results}, robust constraints on the 
C/O ratios and thermal structures of hot Jupiter atmospheres can be placed currently from thermal spectra 
using a combination of existing {\it Spitzer} and ground-based data with new data from {\it HST} WFC3 and additional ground-based observations. The {\it HST} WFC3 bandpass would be critical to constrain the H$_2$O abundances in hot Jupiter atmospheres, especially for planets for which a {\it Spitzer} 5.8 $\micron$ observation is not available. At the same time, ground-based near-infrared observations provide a very powerful means to constrain the temperature structure independent of the composition. The $J$, $H$, and $K$ bands probe deeper in the planetary atmosphere, compared to the mid-IR {\it Spitzer} bands, and hence constrain the temperature structure in the lower atmosphere, which 
is important to constrain a thermal inversion in the upper atmosphere. Atmospheres with thermal inversions 
in the upper-atmospheres tend to have lower temperatures in the lower atmospheres, and vice versa 
(see e.g. Burrows et al. 2008, Madhusudhan \& Seager, 2010). Once the temperature profile is partially 
constrained by the  near-infrared data, the {\it Spitzer} data along with any additional observations in the 
molecular bands, allow good constraints on the molecular composition and hence the C/O ratio. 

For very hot atmospheres, the near-infrared region between 0.8 and 1.1 $\micron$ also provide an important 
diagnostic for the temperature profile as well as the C/O ratio. This wavelength region contains strong spectral 
features due to TiO and VO. As discussed in the case of WASP-19b, in Section~\ref{sec:results}, the $z^\prime$ band, $Y$ band, $z$ band, and part of the $J$ band, all between 0.9 and 1.1 $\micron$, can be useful probes of the TiO abundance and, indirectly the C/O ratio.  If a highly irradiated atmosphere, such as WASP-19b, is known to have 
no thermal inversion, based on longer wavelength data, then one or more of these four bands ($z^\prime$ to $J$) 
can be used to distinguish between a C-rich and an O-rich atmosphere. A highly irradiated O-rich atmosphere 
without a thermal inversion will have strong TiO absorption, and hence low flux, in these bands. Whereas a C-rich 
planet without a thermal inversions can have high flux in these bands, depending on how much TiO and VO are 
depleted due to the high C/O. Observations in the same bands for a model with a thermal inversion are discussed 
for the case of WASP-33b in Section~\ref{sec:results}. 

New observations aiming to determine C/O ratios will benefit most from observing the hottest of 
hot Jupiters. As discussed in Section~3, the contrast in molecular compositions between 
C-rich and O-rich atmospheres occurs increases with temperatures. While the contrast is minimal 
for temperature below $\sim 1200 K$, the contrast is maximal for temperatures above $\sim$ 2000 K.  
Direct elemental estimates are possible if observations can be made in the H$_2$O, CO, and/or CH$_4$ 
bands, and, if possible, in bands of C$_2$H$_2$ and HCN. The inference of C/O ratios is straightforward at high temperatures. A strong H$_2$O feature and/or a weak CH$_4$ feature is an indication of C/O $<$ 1. Conversely, a weak H$_2$O feature and/or a noticeable CH$_4$ feature is an indication of C/O $\ge$ 1. Observations in CO bands should always find a strong CO signature for hot Jupiters, independent of the C/O ratio. The CO/H$_2$O ratio will be unity for 
C/O=0.5, and can be as high as 10-1000 for higher C/O ratios. More detailed inferences based 
on model spectra are discussed in sections~\ref{sec:spectra} and \ref{sec:results}. 

We identify a sample of 20 hot Jupiters which are currently most suitable for measuring C/O ratios, as 
shown in Fig.\ref{fig:irac_systems}. These planets all have $T_{\rm q,f=0} \gtrsim 1300$, and have already 
been observed in at least three channels of {\it Spitzer} photometry, in the cryogenic phase\footnote{
In the post-cryogenic "warm" phase now only two of {\it Spitzer} 's original six photometric channels 
are operational.} The {\it Spitzer} channels cover features of all the prominent molecules: 3.6 $\micron$ 
(H$_2$O and CH$_4$, C$_2$H$_2$, HCN), 4.5 $\micron$ (CO, CO$_2$, C$_2$H$_2$, HCN), 5.8 $\micron$ (CO and H$_2$O), 8 $\micron$ (H$_2$O, CH$_4$, C$_2$H$_2$, HCN), 16 $\micron$ (CO$_2$ and H$_2$O), 
and 24 $\micron$ (H$_2$O). As such, combining existing {\it Spitzer} data for systems shown in 
Fig.\ref{fig:irac_systems} with new near-IR observations with {\it HST} and ground-based facilities 
can lead to C/O estimates for a statistically significant sample of 20 hot Jupiter atmospheres. 

High confidence estimates of C/O ratios can be made either with spectrophotometric observations that 
can resolve molecular bands or by combining several broadband photometric observations over a 
wide wavelength range, or both. Currently available spectrophotometric facilities include the {\it HST} WFC3 
instrument in the $1 - 1.7$ $\micron$ bandpass (Deming et al. 2010 HST Program \#12736), 
ground-based spectrophotometry (e.g. Swain et al. 2010; Mandell et al. 2011; Crossfield et al. 2011; Bean et al. 2011), and SOFIA (Angerhausen et al. 2010), in the near term, and the {\it James Webb Space Telescope} in the future.  On the other hand, photometric observations can be made with {\it Warm Spitzer} at 3.6 $\micron$ and 4.5 $\micron$, and ground-based near-IR photometry in the $z^\prime$ (Lopez-Morales et al. 2010), $J$, $H$, and $Ks$ 
(Croll et al. 2011). 

While the above discussion focused primarily on observations of thermal emission from exoplanets, transmission spectra can also be used to constrain the C/O ratios in giant exoplanetary atmospheres. Since transmission spectra are less sensitive to temperature structure, and hence to thermal inversions, the benefits of observing in continuum such as broadband photometry in $H$ or $K$ bands is minimal, except to measure a reference absorption or an accurate photospheric radius of the planet. However, transmission spectroscopy or narrow-band photometry in the molecular bands would be a powerful means to constrain the chemical composition. Some of the useful bands in that regard would be the {\it HST} WFC3 bandpass between 1.1 and 1.7 $\micron$ for H$_2$O, the 0.8 - 1.1 $\micron$ region for TiO/VO, parts of the $H$ (1.6 $\micron$) and $K$ (2.1 $\micron$) bands for CH$_4$ and other hydrocarbon features, and the CO band around $\sim$ 2.3 $\micron$. If either H$_2$O or CH$_4$ abundance can be measured in transmission, the C/O ratio can be estimated. If only CO is detected, it would not be adequate to determine the CO ratio. Measurement of any hydrocarbon abundance in a hot atmosphere would be a strong evidence of a carbon-rich atmosphere. 

\section{Summary and Conclusions}
\label{sec:discussion}

We propose the C/O ratio as a new dimension for atmospheric characterization of extrasolar planets. 
In the past, atmospheric models for exoplanets have typically assumed solar elemental 
abundances. Such models have been used for one-dimensional 
characterization of irradiated giant planets based primarily on the degree of incident irradiation, or 
equivalently temperature, while fixing the elemental abundances to solar values, 
with C/O = 0.5. However, recent observations of exoplanetary atmospheres 
have revealed extreme departures from predictions based on models and classification schemes 
that assumed solar abundances. In this work, we find that the C/O ratio plays a critical role in 
governing the atmospheric chemistry, and non-solar C/O ratios can potentially explain observations 
which were hitherto reported as anomalous using solar abundance models. We, therefore, 
propose a two-dimensional classification system for exoplanetary atmospheres where chemistry, 
represented by the C/O ratio, and temperature ($T$), or equivalently incident irradiation, form the two 
dimensions.

Our proposed 2-D characterization scheme opens a new physico-chemical phase space in which to interpret  observations of H$_2$-dominated exoplanetary atmospheres. In hot H$_2$-dominated atmospheres, the C/O ratio critically influences the concentrations of major molecular species, such as H$_2$O, CO, and CH$_4$. For $T \gtrsim 1200$ K, a C/O of 1 constitutes a natural boundary between the familiar O-rich (C/O $<$ 1) and the exotic C-rich (C/O $\ge$ 1) atmospheres. Hot O-rich atmospheres are expected to be dominated by H$_2$O and CO, and be less abundant 
in CH$_4$; CH$_4$ decreases as $T$ increases and becomes negligible for $T \gtrsim 1500$ K. On the other hand, C-rich atmospheres at high $T$ are expected to be depleted in H$_2$O, and enhanced in CH$_4$, C$_2$H$_2$, HCN,  and in the CO/H$_2$O ratio, compared to those obtained with a solar C/O of 0.5. Each of the O-rich and C-rich regimes can be further divided into two subclasses depending on the temperature or incident irradiation, analogous to previous classification schemes (Fortney et al. 2008). However, C-rich atmospheres are unlikely to host thermal inversions 
due to TiO/VO even for the most highly irradiated cases, as demonstrated by Madhusudhan et al. (2011b). For C/O $\geq 1$, almost all the oxygen is occupied by CO, which is the dominant O bearing molecule at high $T$, leaving little O for other oxygen-bearing molecules. 

We identify four classes of irradiated atmospheres (O1, O2, C1, and C2) with distinct compositional 
and spectroscopic properties. Classes O1 and O2 are O-rich (C/O $<$ 1), and classes C1 and C2 
are C-rich (C/O $\geq$ 1). The division between O1 and O2, and between C1 and C2 is based on irradiation (or temperature); O1 and C1 are the cooler classes. In O1 and O2 atmospheres, H$_2$O is a major 
oxygen-carrier at all temperatures, and CO or CH$_4$ is the dominant carbon-carrier, depending on the 
temperature. The division between the O1 and O2 classes is analogous to the pL and pM classes, respectively, 
of Fortney et al. (2008). O2 classes might be likely to host thermal inversions due to TiO because of their 
high irradiation levels. On the other hand, C1 and C2 classes are characterized by depleted H$_2$O 
and enhanced CH$_4$, C$_2$H$_2$, and HCN compared to O-rich atmospheres at the same level of irradiation. The differences in H$_2$O and CH$_4$ abundances between O-rich and C-rich atmospheres are enhanced with temperature and can reach upto factors of $10^3$ for $T \gtrsim 2000$ K. In the cooler C1 class, however, the H$_2$O 
depletion is too low to be distinguishable from atmospheres with the same irradiation in the O1 class. 
Consequently, C/O determination of C1 class atmospheres would require high resolution spectroscopy and/or 
other molecular species as tracers of C/O $\geq 1$. Finally, neither C1 nor C2 class atmospheres can host 
thermal inversions due to TiO or VO. 

We emphasize the importance of considering non-solar C/O ratios in interpreting observations of exoplanetary 
atmospheres. We find that non-solar C/O ratios can potentially explain observations of several hot Jupiters which have hitherto been regarded as anomalous in the literature, based on solar abundance models. For example, while observations of one of the coolest hot Jupiters, XO-1b, were interpreted as indication of a thermal inversion (Machalek et al. 2008), some of the hottest hot Jupiters like WASP-12b, WASP-14b, and WASP-19b showed no evidence of a strong thermal inversion (Madhusudhan et al. 2011a). On the other hand, previous studies have been unable to explain observations of CoRoT-2b. We find that the data for both XO-1b and CoRoT-2b, along with several other hot Jupiters can be explained by models with non-solar C/O rations and with different constraints on their temperature structures. These solutions agree with predictions based on the 2-D characterization scheme proposed in this work. The candidate C-rich atmospheres of XO-1b, CoRoT-2b, WASP-19b, and WASP-33b, reported in this work can be confirmed with follow-up observations with existing facilities. 

We present observational strategies to constrain atmospheric C/O ratios of currently known exoplanets with 
existing and forthcoming facilities. The differences between C-rich and O-rich systems increase with temperature. 
As such, the hotter atmospheres are more favorable for obtaining C/O constraints. We identify 20 currently known transiting giant exoplanets which are most suitable for C/O constraints and 2-D characterization. The identified planets all have $T \gtrsim 1200$ K in their lower atmospheres, and have already been observed in at least three channels of {\it Spitzer} IRAC photometry. As demonstrated for the case of WASP-12b (Madhusudhan et al. 2011a),  combining the Spitzer observations with near-infrared observations can provide statistically meaningful joint constraints on C/O ratios and temperature profiles. Near-infrared observations may include ground-based photometry (e.g. in the $z$, $J$, $H$, and $K$ bands; Lopez-Morales et al. 2010; Croll et al. 2011) and spectroscopy (Bean et al. 2010; Mandell et al. 2010; Swain et al. 2010; Crossfiled et al. 2011), and space-borne spectroscopy with HST WFC3 (e.g. Deming et al. 2010 HST Program \#12736) and, potentially, SOFIA (Angerhausen et al. 2010). More stringent constraints will be possible with the {\it James Webb Space Telescope} in the longer term. 

If atmospheric elemental abundances reflect the abundances in planetary interiors, exoplanets in the O-rich and C-rich regimes are expected to span a diverse range of interior compositions and planet formation scenarios.  In particular, C-rich atmospheres initiate the new class of carbon-rich planets (CRPs; Madhusudhan et al. 2011b) with atmospheres, interiors, and formation scenarios markedly different from expectations based on the solar abundances. While O-rich chemistry is expected to form silicates in planetesimals and rocky cores, C-rich environments would be expected to be dominated by carbides (such as SiC) and pure carbon minerals such as graphite and diamond (Gilman 1969; Larimer 1974; Kuchner \& Seager, 2005; Bond et al. 2010). Equally drastic are the protoplanetary conditions required to form CRPs, such as a naturally carbon-rich gas in the disk (Oberg et al. 2011), or a sink of oxygen in the disk (Madhusudhan et al. 2011b), or the predominance of planetesimals rich in carbonaceous material like tar, instead of water-ice (Lodders, 2004). Thus, determination of C/O ratios in exoplanetary atmospheres are critical for understanding their interiors and formation scenarios, which have thus far been commonly studied assuming solar abundances. 

\acknowledgements{The author acknowledges support from the Yale Center for Astronomy and Astrophysics (YCAA), Yale University, through the YCAA postdoctoral fellowship. The author thanks Heather Knutson, Drake Deming, Julianne Moses, Richard Freedman, Adam Burrows, Sara Seager, Jonathan Fortney, Adam Showman, Mark Marley, David Spiegel, Avi Mandell, Jean-Michel Desert, Joe Harrington, Jacob Bean, David Anderson, Jasmina Blecic, Ian Crossfield, and the anonymous referee for helpful comments and/or discussions. The author thanks Mark Swain for sharing their WFC3 data of WASP-12b. This research has made use of the Exoplanet Orbit Database and the Exoplanet Data Explorer at exoplanets.org. \newline}

\end{document}